\documentclass[prb,aps]{revtex4}
\usepackage{amsmath,enumerate}
\usepackage{amssymb}
\usepackage{graphicx}
\usepackage{amsthm}
\newtheorem{theorem}{Theorem}

\newtheorem{lemma}{Lemma}

\newtheorem{prop}{Proposition}

\newcommand{\be}{\begin{equation}}
\newcommand{\ee}{\end{equation}}
\newcommand{\bea}{\begin{eqnarray}}
\newcommand{\eea}{\end{eqnarray}}
\newcommand{\bra}[1]{\big\langle #1\big|}
\newcommand{\ket}[1]{\big|#1\big\rangle}
\newcommand{\braket}[2]{\big\langle #1,#2\big\rangle}
\newcommand{\ketbra}[2]{|#1\rangle\langle #2|}
\newcommand{\pure}[1]{\ketbra{#1}{#1}}
\newcommand{\Tr}{{\mathop{\mathrm{Tr}}}}
\providecommand{\one}{\leavevmode\hbox{\small1\kern-3.8pt\normalsize1}}
\renewcommand\Im{\operatorname{Im}}
\newcommand{\Z}{\mathbb{Z}}

\newcommand{\D}{{\cal D}}
\newcommand{\DX}{{\cal D}_X}
\newcommand{\DY}{{\cal D}_Y}
\newcommand{\gap}{\gamma}
\newcommand{\A}{\mathcal{A}}
\newcommand{\N}{\mathbb{N}}
\newcommand{\R}{\mathbb{R}}
\newcommand{\Sanew}[1]{\mathcal{S}_{#1}}
\newcommand{\Sanewloc}[2]{\mathcal{S}^{(#1)}_{#2}}
\newcommand{\Sa}{\Sanew{\Delta}}
\newcommand{\Saloc}{\Sanew{\Delta,r}}
\newcommand{\diam}{\operatorname{diam}}
\newcommand{\supp}{\operatorname{supp}}
\newcommand{\clr}{C_{qa}}
\newcommand{\qjl}{Q_{\max}\, J \,L\ln^2 L}
\numberwithin{lemma}{section}

\makeatletter
\def\imod#1{\allowbreak\mkern10mu({\operator@font mod}\,\,#1)}
\makeatother

\begin{document}

\title{Quantization of Hall Conductance For Interacting Electrons on a Torus}
\author{Matthew B. Hastings}
\email{mahastin@microsoft.com}
\affiliation{Microsoft Research Station Q, CNSI Building, University of California, Santa Barbara, CA,
93106, USA}
\author{Spyridon Michalakis}
\email{spiros@caltech.edu}
\affiliation{Institute for Quantum Information and Matter, Caltech - Pasadena, CA, 91125, USA}

\begin{abstract}
We consider interacting, charged spins on a torus described by a gapped Hamiltonian with a unique groundstate and conserved local charge. Using quasi-adiabatic evolution of the groundstate around a flux-torus, we prove, without any averaging assumption, that the Hall conductance of the groundstate is quantized in integer
multiples of $e^2/h$, up to exponentially small corrections in the linear size $L$. In addition, we discuss extensions to the fractional quantization case under an additional topological order assumption on the degenerate groundstate subspace.

\end{abstract}
\maketitle

\section{Introduction}
At low temperature, the Hall conductance of a quantum system can be
quantized to remarkable precision.  While this is an experimental fact,
and the essential ingredients of our intuitive understanding of it were
provided by Laughlin~\cite{laughlin}, there is still no fully satisfactory
mathematical proof of why this happens. The main approaches theoretically
are either the Chern number approach~\cite{avron,niu,avron2}, which relies on either
an additional averaging assumption or a uniformity assumption, and the non-commutative geometry approach~\cite{ncg}, which is only applicable to non-interacting electrons. Another approach by Fr\"ohlich and collaborators~\cite{frohlich} uses field-theoretic methods; while that approach uses a gap to prove quantization, it does not work at the level of a microscopic Hamiltonian and so does not give bounds in terms of the underlying constants such as interaction strength and range. In this paper, we present an approach which avoids all these issues.  It avoids averaging and is applicable to interacting electrons; it gives quantitative bounds in terms of the microscopic interaction
constants; and finally it can even be extended to handle fractional quantization.

Before we formulate our main result, we discuss our setup of the Quantum Hall Effect.
We will consider a discrete, tight-binding model of degrees of freedom, which may include fermions, spins, or both,
at sites on a torus $T$, where each site $s$ has a finite dimensional Hilbert space $\mathcal{H}_s$ associated with it (the dimension of the Hilbert space does not in any way enter into our bounds). The torus $T$ is obtained by joining the boundaries of a finite $[1,L]\times [1,L]$ subset of $\Z^2$. Let $\A_Z$ be the algebra of observables on set $Z$ which have even fermionic parity (see Appendix \ref{Hilbert} for a detailed construction  of the Hilbert space of the system and for definition of fermionic parity).
At each site $s\in T$, we introduce the charge operator $q_s \in \A_{\{s\}}$ with eigenvalues $0$, $1, \dots , q_{\max}$, representing a state with the respective charge at site $s$.
From a general point of view, we are interested in properties of the groundstate
of the Hamiltonian $H_0 = \sum_{Z\subset T} \Phi(Z)$, where the following assumptions are satisfied:
\begin{enumerate}
\item The Hamiltonian terms $\Phi(Z)$ are $k$-body interactions with finite strength and finite range.
Formally,  $\exists\,  J > 0$ and $R,k_{\max} \ge 1$ with $L>2R$ such that:
\begin{enumerate}
\item $\Phi(Z) = \Phi^{\dagger}(Z) \in \A_Z$, 
\item $\sup_{s\in T} \sum_{Z\ni s} \| \Phi(Z) \| \leq J$, 
\item $\forall Z\subset T,$ if $\diam(Z) > R \mbox{ or } |Z| > k_{\max}$, then $\Phi(Z) = 0,$
\end{enumerate}
where
$\diam(Z) = \max_{\{ s_1,s_2\in Z\}} \{d(s_1,s_2)\}, \quad d(s_1,s_2) = |x(s_1)-x(s_2)| \imod{L} + |y(s_1)-y(s_2)| \imod{L},$
with $x(s)$, $y(s)$ denoting the $x$ and $y$ coordinates of site $s$, respectively.
\item The Hamiltonian $H_0  = \sum_{Z\subset T} \Phi(Z)$ has a unique,
normalized groundstate denoted by $\ket{\Psi_0}$ and a
spectral gap to the first excited state which is lower bounded by $\gap > 0$. 
We use $P_0=|\Psi_0\rangle \langle \Psi_0|$ to denote the projector onto the groundstate.
\item The total charge $Q = \sum_{s\in T} q_s$ is conserved, so that $[Q,H_0]=0$.
\end{enumerate}

Our results are quantitative, giving error bounds that decay almost exponentially fast as $L \rightarrow \infty$.  
We refer to a function $f$ as ``almost-exponentially decaying", if for all $c$ with $0 \leq c <1$ there is a constant $C$ such that $f(x) \leq C \exp(-x^c)$ for all sufficiently large $x$.  
We say that a quantity is almost-exponentially small in some other quantity $x$, if it is bounded in absolute value by an almost-exponentially decaying function of $x$.
This family of almost-exponentially decaying functions has the nice property that an almost-exponentially decaying function multiplied by a polynomial is still almost-exponentially decaying.  Also, if $f(x)$ is almost-exponentially small in $x$, then $f(ax)$ is almost-exponentially small in $x$ for any $a>0$.  These properties may help the reader; while we have many detailed error bounds, many of them are in this form of an almost-exponentially decaying function (functions $f,g$ defined later) times a polynomial, and using this property one can verify that our final error is almost-exponentially decaying without having to verify all the manipulations of polynomials in the intermediate steps.

Our main result is the following:
\begin{theorem}
For any fixed, $L$-independent $R,J, q_{\max}$ and spectral gap $\gamma>0$, for any Hamiltonian $H_0$ satisfying the above assumptions, the difference between the Hall conductance $\sigma_{xy}$ and the nearest integer multiple of $e^2/h$ is almost-exponentially small in $L$, where the Hall conductance is defined in Eq.~(\ref{def:cond}) and
$e^2/h$ denotes the square of the electron charge divided by Planck's constant.
\end{theorem}
This implies that, for fixed $J,R,\gamma$ and $q_{\max}$, the difference between the Hall conductance and the nearest integer multiple of $e^2/h$
tends to zero in the $L\rightarrow \infty$ limit; we remark that this integer might, however, depend upon $L$.  It also implies that given any continuous path of Hamiltonians $\{H_s\}$ with uniform bounds on $J,R,\gamma$ for all $s \in [0,1]$, then (for sufficiently large $L$) the closest integer to $\sigma_{xy} \cdot (h/e^2)$ is constant along the path. The almost-exponentially small function in the theorem above is a sum of three different quantities computed later in the proof; the largest one (asymptotically) is given in Eq.~(\ref{B3bound}).

In the above theorem, we specify that $R,J,\gamma,q_{max}$ are $L$-independent.  For brevity, 
later in the paper, whenever we write that a quantity is almost-exponentially small in $L$, this will always similarly be done for
fixed, $L$-independent choices of
$R,J,q_{max}$ and $\gamma>0$, even if not explicitly specified.

Before giving the proof, we discuss the applicability of our assumptions to physical experimental systems. The first assumption includes only finite range interactions. While there are long range Coulomb interactions in real
experimental systems, the screening of Coulomb interactions may justify this assumption. Moreover, the case in which the torus is not exactly square, but has an aspect ratio of order unity, can be handled by combining several sites in one direction into a single site to make the aspect ratio unity; in fact, with minor changes in the proof, a polynomial aspect ratio can be handled also, at the cost of a polynomial increase in the error bounds.
We omit this case for simplicity. Our techniques can also be extended to the case of exponentially decaying interactions, which we also omit for the sake of brevity.

We note here that interactions $\Phi(X)$ in $H_0$, supported on subsets $X$ which cross the boundaries $x= 0$ and $y=0$, are non-zero on the torus $T$.  
Furthermore, without loss of generality, we may assume that each term $\Phi(Z)$ of the initial Hamiltonian commutes with the total charge $Q$, by
substituting $\Phi(Z)$ with the averaged interaction $\Phi'(Z)$, which has the same range $R$ and strength still bounded by $J$:
\be \label{defn:interaction}
\Phi'(Z) = \frac{1}{2\pi}\int_0^{2\pi} e^{i \theta Q}\, \Phi(Z)\, e^{-i \theta Q}\, d\theta.
\ee
In particular, note that $e^{i \theta Q}\, \Phi(Z)\, e^{-i \theta Q} = e^{i \theta Q_Z}\, \Phi(Z)\, e^{-i \theta Q_Z}$, where $Q_Z = \sum_{s\in Z} q_s$, $\|\Phi'(Z)\| \le \|\Phi(Z)\|$ and: 
$$[Q,\Phi'(Z)] = \frac{1}{2\pi i}\int_0^{2\pi} \partial_\theta \left(e^{i \theta Q}\, \Phi(Z)\, e^{-i \theta Q}\right)\, d\theta = \frac{1}{2\pi i} \left(e^{i 2\pi Q}\, \Phi(Z)\, e^{-i 2\pi Q} - \Phi(Z)\right) = 0,$$ since the total charge $Q$ is the sum of commuting charges $q_s$ with integer spectrum. Moreover, $$\sum_{Z\subset T} \Phi'(Z) =  \frac{1}{2\pi}\int_0^{2\pi} e^{i \theta Q}\, H_0 e^{-i \theta Q}\, d\theta = H_0,$$ since $H_0 = \sum_{Z\subset T} \Phi(Z)$ commutes with the total charge $Q$. Hence, when we write $\Phi(Z)$ from now on, we will be referring to $\Phi'(Z)$ defined in (\ref{defn:interaction}). 
Given the motivation behind the proof that follows, it is important to note that the above transformation is not an added assumption on $H_0$, but a consequence of the assumptions already enumerated. The {\it averaging assumption} that has persisted in all previous attempts to prove quantization of the Hall conductance for interacting systems is of a different nature and refers to the adiabatic transformation of the Hamiltonian $H_0$ into a family of distinct Hamiltonians, as discussed below. In contrast, the above {\it symmetrizing transformation} is a regrouping of the interaction terms in $H_0$.  We remark that if the terms $\Phi(Z)$ in $H_0$ had individually commuted with $Q$ before the symmetrizing transformation, then the symmetrizing transformation has no effect, giving $\Phi'(Z)=\Phi(Z)$.

\subsection*{Sketch of the main argument.}
Our proof is closely related to the Chern number approach \cite{avron,niu}, so we review it and then contrast our technique.
In the Chern number approach, one computes the Hall conductance of the above system of interacting electrons by introducing magnetic fluxes through twists $\theta_x, \theta_y$ at the boundary of the torus, $T$.  A time-dependent flux $\theta_x$ ``generates'' an electric field in the $x$ direction, while the flux $\theta_y$ ``measures'' the current in the $y$ direction, and thus the Hall conductance can be identified with a certain current-current correlation evaluated at the groundstate $\ket{\Psi_0}$.  In the Chern number approach, one assumes that there is a unique groundstate for all $\theta_x,\theta_y$.  
If the groundstate is adiabatically transported around an infinitesimal loop near $\theta_x=\theta_y=0$, the state acquires a phase proportional to the area of the loop multiplied by the Hall conductance. That is, the Hall conductance is equal to the curvature of the connection given by parallel transport of
the groundstate. The average of the Hall conductance over the torus is then equal to an integer, the Chern number. Unfortunately, this argument relies on unknown spectral properties of the Hamiltonians involved in the adiabatic evolution, so it can only provide a quantization result for an averaged Hall conductance, which may or may not correspond to the Hall conductance at the groundstate of $H_0$.

Our proof requires only the presence of a spectral gap for $H_0$, which allows us to construct a simulated adiabatic evolution, introduced in Ref.~\onlinecite{hast-lsm} as {\it quasi-adiabatic evolution} (and studied further in Ref.~\onlinecite{osborne,hast-quasi, BMNS:2011}.) There are two crucial reasons for using the quasi-adiabatic evolution in its latest incarnation \cite{hast-quasi, BMNS:2011}: 
\begin{enumerate}
\item The action of the quasi-adiabatic evolution on the groundstate subspace is indistinguishable from the action of the adiabatic evolution (cf. Kato's adiabatic theorem~\cite{Kato}, originally formulated by M. Born and V. Fock in 1928), as long as there is a non-zero, constant lower-bound on the spectral gap throughout the evolution. This indistinguishability between the two evolutions implies that the geometric phase acquired by the groundstate $\ket{\Psi_0}$ through quasi-adiabatic evolution on an infinitesimal loop near the origin in flux-space, will be the desired Hall conductance multiplied by the area of the loop, as explained at the beginning of this section.
\item The quasi-adiabatic evolution is generated by quasi-local interactions with well defined decay and strength, precisely the ingredients one needs to define an effective light-cone for the propagation of correlations - the kind of correlations responsible for the quantization of the Hall conductance. In particular, quasi-adiabatic evolution has the added benefit of generating an evolution in flux-space where the parallel transport of the groundstate has {\it uniform curvature}, up to exponentially small errors in the linear size of the system.  The locality of the evolution generated by quasi-adiabatic continuation holds even in regions of flux-space where the spectral gap of the Hamiltonian may vanish; in these regions, the quasi-adiabatic evolution may differ from Kato's adiabatic evolution, in particular, by accessing higher energy states during intermediate steps of the evolution.
\end{enumerate}

Our proof rests on four results.
First, in Proposition~\ref{prop:adiabatic_phase_3} proven in Section~\ref{sec:loop_unitaries}, we show that the quasi-adiabatic evolution of the groundstate around a small (but non-infinitesimal) square loop near the origin in parameter space ($\theta_x=\theta_y=0$) gives back the initial groundstate up to a geometric phase, which we relate to the Hall conductance at $\theta_x=\theta_y=0$. This result uses the indistinguishability of quasi-adiabatic and adiabatic evolution given a uniform lower bound on the spectral gap. By taking the loop size sufficiently small, the existence of a gap at the origin implies the existence of a gap along the loop.  In contrast, the later steps of the proof use loops that do not remain near the origin and, crucially, in these steps of the proof no gap along the path is needed or assumed. The size of the small loops is chosen in~\eqref{rchoice} to be almost-exponentially small in the linear size $L$, in order to optimize the final error bound.

Second,  we show in Proposition~\ref{prop:gs_evol} that quasi-adiabatic evolution of the groundstate around a large loop in parameter space,
$\Lambda: (0,0) \rightarrow (2\pi,0) \rightarrow (2\pi,2\pi) \rightarrow (0, 2\pi) \rightarrow (0,0),$  returns to a state which is almost-exponentially close to the initial groundstate with a {\it trivial overall phase}. To do this, we use locality estimates and energy estimates to prove that quasi-adiabatic evolution of the ground state along a single side of this path, such as  $(0,0) \rightarrow (2\pi,0)$, returns to a state almost exponentially close to the ground state. Then, we use a cancellation of phases between opposite sides of the path to show that the overall phase is trivial.

\begin{figure}
\centering
\includegraphics[width=160px]{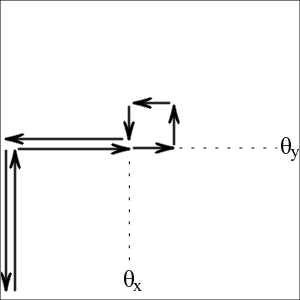}
\caption{{\small{Arrows illustrating the path in flux-space that the quasi-adiabatic evolution follows in lemma \ref{lem:translation}, where it is shown that the state at the end of the evolution is the same as the initial groundstate, up to a geometric phase that depends only on the size of the small loop. In particular, the phase is independent of $(\theta_x,\theta_y)$, up to vanishing corrections in the lattice size, $L$.
}}}
\label{fig:flux_path}
\end{figure}

Third, we show, in a procedure reminiscent of Stokes' theorem, that the quasi-adiabatic evolution around the large loop in parameter space considered in the second step
can be decomposed into the product of a large number of quasi-adiabatic evolutions around paths as illustrated in Fig.~\ref{fig:flux_path} (see also, Fig.~\ref{fig:decomposition}). Each of these paths consists of following a path in flux-space starting at $(0,0)$, going to $(0,\theta_y)$, then to $(\theta_x,\theta_y)$, evolving counter-clockwise around a small loop at the given $(\theta_x,\theta_y)$ and then returning to $(0,0)$, as shown in Fig. \ref{fig:flux_path}. That is, we decompose the motion around a single large loop into many motions around small loops with different basepoints.

In the fourth step, we show that quasi-adiabatic evolution of the groundstate following the paths described in the third step gives a state almost-exponentially close to the initial groundstate, up to a geometric phase. Crucially, we prove in lemma \ref{lem:translation} that the geometric phase is independent of $\theta_x,\theta_y$, up to almost-exponentially small errors in the system size. \footnote{Note, however, that this does {\it not} imply that the curvature of adiabatic evolution tends to a constant for large $L$, independent of $\theta_x,\theta_y$, as adiabatic and quasi-adiabatic evolution may differ in general if the gap becomes small away from $\theta_x=\theta_y=0$.} Thus, the first step in the proof implies that the geometric phase generated by any one of these evolutions corresponds to the Hall conductance at $\theta_x = \theta_y = 0$, multiplied by the area of the small loop.  As in the second step, we do {\it not} require any lower bound on the spectral gap of the Hamiltonian path in this step, and if the gap does become small, then the quasi-adiabatic evolution around this path may not simulate the true adiabatic evolution. Rather, the proof that we return to the groundstate, with the right geometric phase, depends on locality estimates based on Lieb-Robinson bounds for quasi-adiabatic evolutions.
Moreover, it is shown in Appendices that this decomposition, combined with the result from the second step of the proof, implies that the phase generated by the evolution around the large loop is the product of the phases coming from the small loops, up to almost-exponentially small corrections.  The bounds on these corrections are in Proposition \ref{prop:stokes}.

Combining these four steps, the trivial phase produced by quasi-adiabatic evolution around the large loop is almost-exponentially close to the product of the phases around each small loop. Since the phase produced by evolution around each small loop is the same (up to almost-exponentially small fluctuations), we can then show that the phase around a small loop near the origin, raised to a power equal to the number of small loops, is close to unity. This geometric phase is proportional to the Hall conductance and the desired quantization follows.

The most difficult technical step is the fourth one. The first step relies only on the fact that quasi-adiabatic continuation for systems with a non-zero spectral gap simulates adiabatic evolution faithfully. The second step is similar to the proof of the Lieb-Schultz-Mattis theorem in higher dimensions~\cite{hast-lsm}.  We use energy estimates and additional virtual fluxes $\phi_x=-\theta_x,\phi_y=-\theta_y$ introduced at $x= L/2+1$ and $y=L/2+1$ to show that the quasi-adiabatic evolution of the groundstate along any of the four sides of the square  (for example, from $\theta_x=\theta_y=0$ to $\theta_x=2\pi,\theta_y=0$) gives a state which is close to the groundstate up to a phase. Then, we show that these phases cancel between evolutions along opposite sides of the flux-square, to produce a trivial overall phase.
We recommend that the reader be familiar with the proof of the higher-dimensional Lieb-Schultz-Mattis theorem before reading this proof, given that some of the ideas of the present proof appear in that proof in a simpler setting.

Lieb-Robinson bounds play a critical role in our proof. These bounds were first introduced in Ref.~\onlinecite{lr}, and extended in Ref.~\onlinecite{hast-koma,ns}. Nachtergaele and Sims, in Ref.~\onlinecite{ns}, gave the important extension to general lattices including exponentially decaying interactions. The most recent, tightest, and most general estimates, which we use, are from \cite{loc-estimates}. The need to use Lieb-Robinson bounds does currently limit us to lattice Hamiltonians; the extension to fermions moving in $\mathbb{R}^2$ would require using unbounded interactions and an infinite dimensional Hilbert space and is currently beyond our techniques, although work such as \cite{anh} is an important step towards this.

We use $C$ throughout the text to refer to various numeric constants of order unity.
There are several places in the proof where we assume some lower bound on $L$ without always explicitly stating it; we do this for three reasons.
The first reason is
that in many places, such as Eq.~(\ref{sets_omega}), we construct subsets of the lattice whose size
is a fraction of $L$; this first reason in fact only requires that $L$ be at least a constant times the range of interactions, $R$.
The second reason is that many of the error estimates are a sum of error terms which have different behaviors in $L$, and we want to require in each case that the term in the error estimate which is dominant in the asymptotic ($L\rightarrow \infty$) limit is larger than the other error terms.  This second reason also only requires that $L$ be at least a constant times $R$.  The third reason is that Proposition~\ref{prop:adiabatic_phase_3} requires an upper bound on $r$. Since the parameter $r$ depends on $L$ as given in Eq.~(\ref{rchoice}), we get a more stringent requirement on $L$ which takes into account implicit units of length set by quantities such as $q_{\max}\,J/\gamma$. After giving the proof, we briefly discuss various extensions. 

\section{Twisted boundary conditions}
Though we are ultimately interested only in computing the Hall Conductance of the groundstate of the Hamiltonian $H_0$ discussed in the previous section, to aid us in our goal we will construct a family of Hamiltonians of the form:
\begin{equation}\label{def:H_S}
H(\theta_x, \phi_x, \theta_y, \phi_y)=\sum_{Z\subset T} \Phi(Z;\theta_x, \phi_x,\theta_y,\phi_y).
\end{equation}

In order to specify the interaction terms in the above family of Hamiltonians, we introduce the following periodic flux-twists: 
For an operator $A$, 
\begin{align}\label{def:charges}
&R_X (\theta_x, A) = e^{i \theta_x Q_X} A e^{-i \theta_x Q_X} = R_X (\theta_x+2\pi, A), &
&Q_X = \sum_{1\le x(s)\leq L/2} q_s&\\
&R_Y (\theta_y, A) = e^{i \theta_y Q_Y} A e^{-i \theta_y Q_Y} = R_Y (\theta_y+2\pi, A), &
&Q_Y = \sum_{1\le y(s)\leq L/2} q_s.&
\end{align}

Then, the interaction term $\Phi(Z;\theta_x, \phi_x,\theta_y, \phi_y)$ is defined according to the following prescription:
\begin{enumerate}[{X}-1.]
\item If $\exists s \in Z : |x(s) - 1| < R$, then $\Phi(Z;\theta_x, \phi_x, \theta_y, \phi_y) = R_X(\theta_x, \Phi(Z;0,0,\theta_y,\phi_y))$.
\item If $\exists s \in Z : |x(s) - (L/2+1)| < R$, then $\Phi(Z;\theta_x, \phi_x,\theta_y, \phi_y) = R_X(-\phi_x, \Phi(Z;0,0,\theta_y,\phi_y))$.
\item Otherwise, $\Phi(Z;\theta_x, \phi_x,\theta_y, \phi_y) = \Phi(Z;0,0,\theta_y,\phi_y)$.
\end{enumerate}
Continuing, the terms $\Phi(Z;0,0,\theta_y,\phi_y)$ are defined as follows:
\begin{enumerate}[{Y}-1.]
\item If $\exists s \in Z : |y(s) - 1| < R$, then $\Phi(Z;0,0,\theta_y, \phi_y) = R_Y(\theta_y, \Phi(Z))$.
\item If $\exists s \in Z : |y(s) - (L/2+1)| < R$, then $\Phi(Z;0,0,\theta_y, \phi_y) = R_Y(-\phi_y, \Phi(Z))$.
\item Otherwise, $\Phi(Z;0,0,\theta_y, \phi_y) = \Phi(Z)$.
\end{enumerate}

\begin{figure}
\centering
\includegraphics[width=150px]{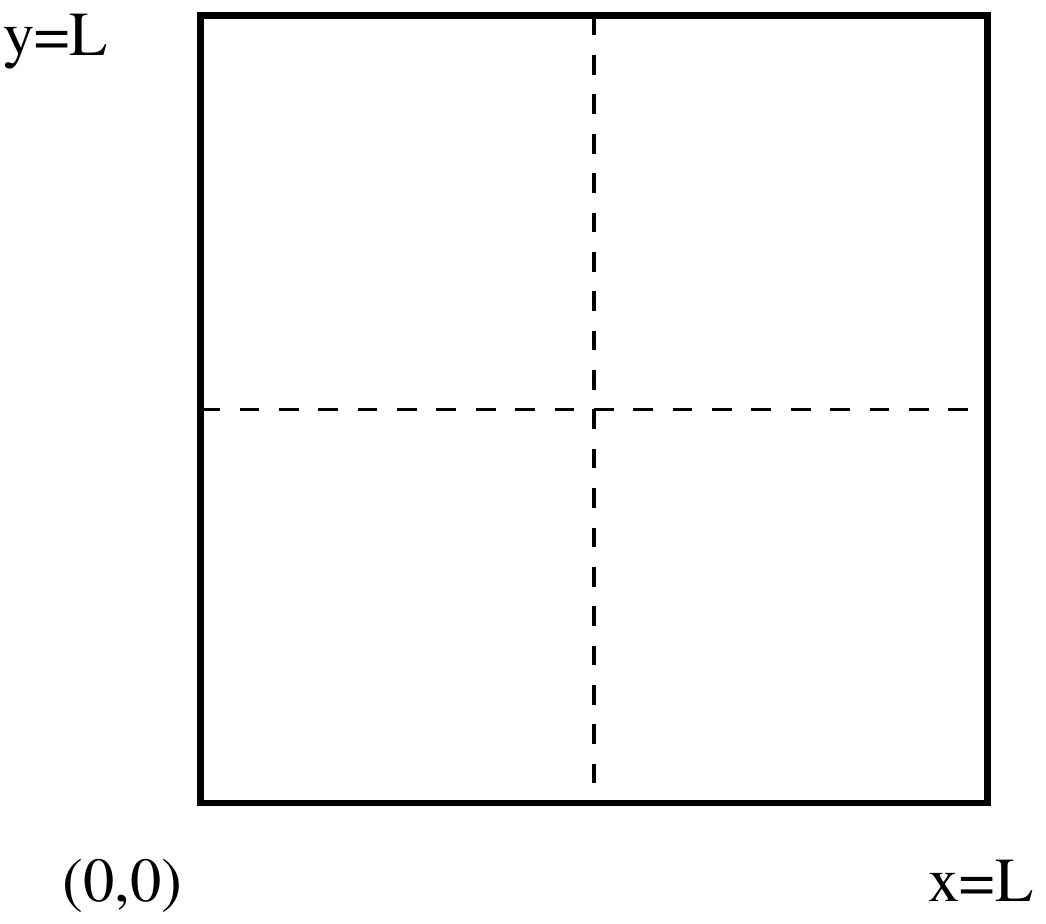}
\caption{{\small{Lines illustrating how the flux-twists change Hamiltonian terms on the torus.  The fluxes $\theta_x,\theta_y$ affect interactions supported on strips of width $2R$ centered on the vertical and horizontal solid lines, while the virtual fluxes $\phi_x,\phi_y$ affect interactions supported on strips of width $2R$ centered on the vertical and horizontal dashed lines. The bulk of interactions remains unchanged.
}}}
\label{fig:interactions}
\end{figure}

Note that since the interactions $\Phi(Z)$ have finite range $R$ and the charge operators $Q_X, Q_Y$ are made up of non-interacting point-charges, the twists only affect two horizontal and two vertical strips, each of width $2R$, centered on the lines $x=1, x = L/2+1$ and $y=1, y= L/2+1$, respectively, as shown in Fig.~\ref{fig:interactions}.
Moreover, for each $\Phi(Z)$, at most two flux-twists act non-trivially. In particular, there are $4$ sets of interactions that see both twists simultaneously, namely those that satisfy conditions X-$1$ $\wedge$ Y-$1$, X-$1$ $\wedge$ Y-$2$, X-$2$ $\wedge$ Y-$1$, or X-$2$ $\wedge$ Y-$2$. For example, interactions $\Phi(Z)$ that satisfy conditions X-$1$ $\wedge$ Y-$1$ transform into $R_Y(\theta_y, R_X(\theta_x,\Phi(Z)))$ (the order in which we apply the twists is irrelevant, since $[Q_X, Q_Y] = 0$).

It will be useful later on to bound the partial derivatives of the Hamiltonian with respect to the flux-twists.
There are $2RL$ sites on each of the strips centered on $x=1$ and $y=1$, on which the individual terms in $\partial_{\theta_x} H(\theta_x,0,\theta_y,\phi_y)$ and $\partial_{\theta_y} H(\theta_x,\phi_x,\theta_y,0)$ may be grouped to act. Moreover, since each interaction $\Phi(Z;\theta_x,0,\theta_y,\phi_y)$ commutes with the total charge $Q$ (see~\eqref{defn:interaction}), we have:
\be
\partial_{\theta_x} \Phi(Z;\theta_x,0,\theta_y,\phi_y) = i\, [Q_{Z\cap X}, \Phi(Z;\theta_x,0,\theta_y,\phi_y)] = -i\, [Q_{Z\cap \overline{X}}, \Phi(Z;\theta_x,0,\theta_y,\phi_y)],
\ee
where $Q_{Z\cap X}$ is the charge in $Z$ that is also contained in $Q_X$ and $Q_{Z\cap \overline{X}} = Q-Q_{Z\cap X}$. Then, each set of interactions acting on a particular site $s$ has norm bounded as follows:
\begin{equation}\label{bound:terms}
\sum_{Z\ni s} \|\partial_{\theta_x} \Phi(Z;\theta_x,0,\theta_y,\phi_y)\| = \sum_{Z\ni s} \|[Q_{Z\cap X}, \Phi(Z;\theta_x,0,\theta_y,\phi_y)]\| \le \min\{|Z\cap X|,|Z\cap\overline{X}|\} \,q_{\max}\, \sum_{Z\ni s} \|\Phi(Z)\| \le Q_{\max}\,\frac{J}{2R},
\end{equation}
where we defined:
\be\label{defn:Q_max}
Q_{\max} \equiv R\, k_{\max}\, q_{\max}.
\ee
We also used that $A \ge 0 \implies \|[A,B]\| = \|[A- (\|A\|/2)\one,B]\| \le 2 \|A- (\|A\|/2)\one\| \|B\| = \|A\|\|B\|$ and $\min\{|Z\cap X|,|Z\cap \overline{X}|\} \le k_{\max}/2$. Similarly, $\sum_{Z \ni s}\|\partial_{\theta} \Phi(Z;\theta_x,\phi_x,\theta,0)|_{\theta=\theta_y}\| \le Q_{\max}\, J/(2R)$. The previous bounds imply:
\begin{eqnarray}\label{bound:rot-H_X}
\|\partial_{\theta} H(\theta,0,\theta_y,\phi_y)|_{\theta=\theta_x}\| &\le& Q_{\max} J L, \\
\|\partial_{\theta} H(\theta_x,\phi_x,\theta,0)|_{\theta=\theta_y}\| &\le& Q_{\max} J L.\label{bound:rot-H_Y}
\end{eqnarray}

Finally, remembering that $\Phi(Z)$ commutes with the total charge $Q$, one may verify using the definitions of twisted interactions, that:
\begin{eqnarray}\label{eq:unitary_equiv_X}
R_X(-\theta, H(\theta_x,\phi_x,\theta_y,\phi_y)) = H(\theta_x-\theta,\phi_x+\theta,\theta_y,\phi_y),\\
R_Y(-\theta, H(\theta_x,\phi_x,\theta_y,\phi_y)) = H(\theta_x,\phi_x,\theta_y-\theta,\phi_y+\theta),\label{eq:unitary_equiv_Y}
\end{eqnarray}
which implies that the twist (near origin)/anti-twist (near midpoint) Hamiltonians (i.e. $\phi_x  = -\theta_x$ and $\phi_y = -\theta_y$) are unitarily equivalent to the original gapped Hamiltonian. In particular, we have the important unitary equivalence between $H_0$ and $H(\theta_x, -\theta_x, \theta_y, -\theta_y)$:
\begin{equation}\label{twist-anti-twist}
R_X(\theta_x, R_Y(\theta_y, H_0)) = H(\theta_x, -\theta_x, \theta_y, -\theta_y).
\end{equation}

\section{The quasi-adiabatic evolution}
A key ingredient of adiabatic evolution is the persistence of a non-vanishing spectral gap. The quasi-adiabatic evolution defined in~\cite{hast-quasi,BMNS:2011} simulates the true adiabatic evolution exactly, so long as the spectral gap is sufficiently large, which is the statement of Lemma~\ref{lem:quasi}. But, first, let us look at some properties of the true adiabatic evolution.
\subsection{Generating the true adiabatic evolution}
In order to simulate the adiabatic evolution of the groundstate of a family of gapped Hamiltonians, we first need to write down a differential equation describing the desired evolution.
To do this, we note that if for some $s \ge 0$, we have that $\{H(\theta)\}_{\theta \in [0,s]}$ is a differentiable family of gapped Hamiltonians with a differentiable family of unique (up to a phase) groundstates $\ket{\Psi_0(\theta)}$ and groundstate energies $E_0(\theta)$, then differentiating $(H(\theta)-E_0(\theta))\,\ket{\Psi_0(\theta)}=0$ yields:
\begin{equation}\label{gs:evol_0}
(\one - P_0(\theta))\partial_{\theta} \ket{\Psi_0(\theta)} = -\frac{\one-P_0(\theta)}{H(\theta)-E_0(\theta)}\, \partial_\theta \, H(\theta) \, \ket{\Psi_0(\theta)},
\end{equation}
where $P_0(\theta) = \pure{\Psi_0(\theta)}$.
Since the phase of $\ket{\Psi_0(\theta)}$ is arbitrary, we may choose:
\be
\ket{\Psi'_0(\theta)} = e^{-\int_0^{\theta} \braket{\Psi_0(t)}{\partial_t \Psi_0(t)} dt}\ket{\Psi_0(\theta)},
\ee
which is also a groundstate of $H(\theta)$ and satisfies the {\it parallel transport} condition:
\be\label{parallel-transport}
\braket{\Psi'_0(\theta)}{\partial_{\theta} \Psi'_0(\theta)} =0.
\ee
Combining the above condition with (\ref{gs:evol_0}) we have:
\begin{equation}\label{gs:evol}
\partial_{\theta} \ket{\Psi'_0(\theta)} = -\frac{\one-P_0(\theta)}{H(\theta)-E_0(\theta)}\, \partial_\theta \, H(\theta) \, \ket{\Psi'_0(\theta)}.
\end{equation}
From here on, we will assume that whenever we {\it adiabatically} transport a unique groundstate along a path in parameter
space, we do this satisfying the {\it parallel transport} condition. Moreover, if $H(\theta)$ has a spectral gap $\Delta(\theta) > 0$, then we have $\Delta(\theta) \cdot (\one - P_0(\theta)) \le H(\theta) - E_0(\theta)$, which combined with (\ref{gs:evol}) implies the bound:
\be\label{bound:partial}
\|\partial_{\theta} \ket{\Psi'_0(\theta)}\| \le \frac{\|\partial_\theta H(\theta)\|}{\Delta(\theta)} 
\ee
Furthermore, after expanding $\partial^2_\theta [(H(\theta)-E_0(\theta)) \ket{\Psi'_0(\theta)}] = 0$ and rearranging terms, we get the higher-order bound:
\begin{eqnarray}
(\one-P_0(\theta))\, \partial^2_\theta \ket{\Psi'_0(\theta)} &=& 
-\frac{\one-P_0(\theta)}{H(\theta)-E_0(\theta)}\, [2\partial_\theta \, H(\theta) \, (\partial_\theta \ket{\Psi'_0(\theta)})+ (\partial^2_\theta H(\theta)) \ket{\Psi'_0(\theta)}] \implies \nonumber\\
\partial^2_\theta \ket{\Psi'_0(\theta)} + \|\partial_{\theta} \ket{\Psi'_0(\theta)}\|^2 \ket{\Psi'_0(\theta)} &=&
-\frac{\one-P_0(\theta)}{H(\theta)-E_0(\theta)}\, [2\partial_\theta \, H(\theta) \, (\partial_\theta \ket{\Psi'_0(\theta)}) + (\partial^2_\theta H(\theta)) \ket{\Psi'_0(\theta)}]\implies \nonumber\\
\|\partial^2_\theta \Psi'_0(\theta)\| &\le& \|\partial_\theta \Psi'_0(\theta)\|^2 + \frac{2\|\partial_\theta \, H(\theta)\| \cdot \|\partial_\theta \Psi'_0(\theta)\| + \|\partial^2_\theta H(\theta)\|}{\Delta(\theta)} \nonumber\\
&\le& \frac{3\|\partial_\theta \, H(\theta)\|^2}{\Delta^2(\theta)} + \frac{\|\partial^2_\theta H(\theta)\|}{\Delta(\theta)}\label{bound:partial_2}
\end{eqnarray}
where we used condition (\ref{parallel-transport}) to get $\braket{\Psi'_0(\theta)}{\partial^2_\theta\Psi'_0(\theta)}= -\|\partial_{\theta} \ket{\Psi'_0(\theta)}\|^2$ from $\partial_\theta\braket{\Psi'_0(\theta)}{\partial_\theta\Psi'_0(\theta)} = 0$, for the second line and used (\ref{bound:partial}) to get the final estimate.

It will be useful to have an estimate on the gap $\Delta(r)$ as $r$ changes from $0$. 
\begin{lemma}\label{lem:gap-estimate}
If for some $s \ge 0$, we have that $\{H(r)\}_{r \in [0,s]}$ is a differentiable path of Hamiltonians each with spectral gap $\Delta(r)$ and groundstate $\Psi_0(r)$, then for $r\in[0,s]$:
\be\label{bound:gap-estimate}
\Delta(r) \ge \Delta(0) - 2\,r \cdot \sup_{\theta \in[0,r]} \|\partial_{\theta} H(\theta)\|
\ee
\begin{proof}
Let $E_0(r)$ be the groundstate energy of $H(r)$.
The spectrum of $H(0)$ is contained in $\{E_0(0)\}\cup [E_0(0)+\Delta(0),\infty)$.  
Therefore, a well-known result in functional analysis (basically, a triangle inequality) implies that the spectrum of $H(r)$ is contained in
$[E_0(0)-\|H(r)-H(0)\|,E_0(0)+\|H(r)-H(0)\|]\cup [E_0(0)+\Delta(0)-\|H(r)-H(0)\|,\infty)$.
We have $\| H(r)-H(0) \| \leq \int_0^r \| \partial_{\theta} H(\theta) \| d\theta \leq r \cdot \sup_{\theta\in[0,r]}\|\partial_{\theta} H(\theta)\|$.
Thus, $\Delta(r) \ge \Delta(0) - 2\,r \cdot \sup_{\theta \in[0,r]} \|\partial_{\theta} H(\theta)\|$.
\end{proof}
\end{lemma}

\subsection{Simulating the true adiabatic evolution.}
We introduce the following super-operator that will allow us to simulate the adiabatic evolution given by (\ref{gs:evol}):
\begin{equation}
\Sa(H,A)=\int_{-\infty}^{\infty} dt\, W_{\Delta}(t)\, \tau_t^{H}(A),
\end{equation}
where we define $\tau_t^{H}(A) := e^{itH} A e^{-itH}$ and $W_{\Delta}(t)$ is the filter function studied in~\cite{hast-quasi,BMNS:2011} with the following properties:
\begin{enumerate}
\item The Fourier Transform satisfies $\hat{W}_{\Delta}(\lambda)=i/\lambda$, for $|\lambda|\ge \Delta$,
\item $W_{\Delta}(t)$ is continuous everywhere but at $t=0$, where it is right continuous with $\|W_\Delta\|_\infty = W_\Delta(0) = 1/2$.  $W_{\Delta}(t)$ is decaying monotonically for $t \ge 0$, with asymptotic decay of order $\exp\{-c_0 \Delta |t|/ \ln^2(\Delta |t|)\}$, for some $c_0 > 0$.
\item $W_{\Delta}(t)$ is an odd function everywhere but at $t=0$. Moreover, $\|W_\Delta\|_1 \le K/\Delta$, for a constant $K > 0$.
\end{enumerate}
Note that the following useful bound holds:
\be \label{naive_bnd:sa}
\|\Sa(H,A)\| \le 2 \int_0^{\infty} W_{\Delta}(t)\, \big\|\tau_{t}^H(A)\big\|\, dt
= 2 \|A\|\, \int_{0}^{\infty} W_{\Delta}(t)\,dt \le (K/\Delta) \, \|A\|
\ee
and that, for Hermitian $H$ and $A$, the operator $\Sa(H,A)$ is also Hermitian.
More importantly, the operator $\Sa(H,A)$ has a decomposition into almost-exponentially decaying interactions~\cite{hast-quasi,BMNS:2011} localized near the support of $A$:
\begin{lemma}\label{lem:local_decomposition}
Define $b_u(r) = \{s \in \Lambda: d(s,u) \le r\}$ to be the ball of radius $r$ centered at $u \in T$. Let $A$ be an operator with $\supp(A) := Z = b_u(r_0)$, $r_0 \ge 0$. Then, setting $Z(M) = \{s\in T: d(s,Z) \le M\}$ and $H_{Z(M)} = \sum_{Y\subset Z(M)} \Phi(Y)$, we have the following decompositions:
\bea
\Sa(H,A) &=& \sum_{r\ge r_0} \Saloc(H,A), \quad \supp(\Saloc(H,A)) = b_u(r), \quad \|\Saloc(H,A)\| \le 2\, \|A\| \, f_\Delta(r-r_0),\label{def:adiabatic_generator}\\
\Sanewloc{M}{\Delta}(H,A) &:=& \sum_{r\ge r_0}^M \Saloc(H,A) = \Sa(H_{Z(M)}, A), \quad \|\Sanewloc{M}{\Delta}(H,A)\| \le (K/\Delta)\, \|A\|,\label{def:local_generator}
\eea
for an almost-exponentially decaying function $f_\Delta(r) \sim \Delta^{-1} \, \exp\{-c_1\, r/\ln^2 r\}$ with $f_\Delta(0)=K/\Delta$ and decay rate $c_1\sim \frac{\Delta}{J\, R}$.
\end{lemma}

We will make extensive use of the localized version $\Sanewloc{M}{\Delta}(H,A)$ of $\Sa(H,A)$, since the new operator has support strictly within $b_u(M)$, for $r_0 \le M \le L$. The following lemma gives an estimate on the error of approximating $\Sa(H,A)$ by $\Sanewloc{M}{\Delta}(H,A)$ for the case $r_0=R$ which will be used later:
\begin{lemma}\label{lemma:local_generator}
For an operator $A$ supported on $b_u(R)$, with $u \in T$, the following bound holds for $M \ge R$:
\be
\|\Sanewloc{M}{\Delta}(H,A) - \Sa(H,A)\| \le \|A\|\, g_{\Delta}(M), \quad g_{\Delta}(M) := 2\sum_{r \ge M} f_{\Delta}(r-R).\label{bnd:local_generator}
\ee
Note that $g_{\Delta}(M) \sim \Delta^{-1} \ln^2 (M-R) \cdot \exp\{-c_1\, (M-R)/ \ln^2 (M-R)\}$, which implies an almost-exponential decay in the error of approximation between the operator $\Sa(H,A)$ and its localized version $\Sanewloc{M}{\Delta}(H,A)$.
\end{lemma}
Remark: The functions $f_\Delta$ and $g_\Delta$ are dimensionful, having dimensions of $({\rm Energy})^{-1}$ (in particular, $1/\Delta$).

It is important to note here that for a Hamiltonian $H(\theta) = \sum_{Z\subset T} \Phi(Z;\theta)$, with each $\Phi(Z;\theta)$ term supported on $Z\subset T$, we have:
\be\label{def:local_generator_H}
\Sanewloc{M}{\Delta}(H(\theta),\partial_\theta H(\theta)) := \sum_{Z\subset T} \Sanewloc{M}{\Delta}(H_{Z(M)}(\theta),\partial_\theta \Phi(Z;\theta)).
\ee
Moreover, the parameter $\Delta$ corresponds to a threshold for the spectral gap of the adiabatic evolution below which the quasi-adiabatic evolution seizes to be faithful to the true adiabatic evolution. More precisely, let us define the following family of states corresponding to a differentiable family of Hamiltonians $H(\theta)$:
\begin{equation}\label{eq:partial_psi}
\partial_{\theta} \ket{\Phi_{\Delta}(\theta)}= i\, \Sa(H(\theta),\partial_{\theta} H(\theta)) \, \ket{\Phi_{\Delta}(\theta)},\quad \ket{\Phi_{\Delta}(0)} = \ket{\Psi_0(0)},
\end{equation}
where $\Psi_0(0)$ is the unique groundstate of $H(0)$.
Defining the unitary $U_\Delta(\theta)$ by its generating dynamics according to the following differential equation:
\begin{equation}\label{eq:dynamics}
\partial_{\theta} U_\Delta(\theta) = i\, \Sa(H(\theta),\partial_\theta H(\theta)) \, U_\Delta(\theta), \, U_\Delta(0) = \one,
\end{equation}
we may also write:
\begin{equation}\label{eq:phi_delta}
\ket{\Phi_{\Delta}(\theta)}= U_\Delta(\theta) \ket{\Psi_0(0)}.
\end{equation}
We say that the state $\Phi_{\Delta}(\theta)$ is the {\it quasi-adiabatic evolution} of $\Psi_{0}(0)$ along the path $0 \rightarrow \theta$.

The following lemma makes precise the sense in which the family of states defined by~(\ref{eq:phi_delta}) are indistinguishable from the true groundstates of the family of Hamiltonians $H(\theta)$:
\begin{lemma}[Adiabatic simulation]\label{lem:quasi} Assume that for some $s \ge 0$, $\{H(\theta)\}_{\theta \in [0,s]}$ is a differentiable path of gapped Hamiltonians, each with spectral gap $\Delta(\theta) \ge\Delta > 0$. For $\ket{\Phi_{\Delta}(\theta)}$ defined in (\ref{eq:partial_psi}) and $\ket{\Psi_0(\theta)}$ the groundstate of $H(\theta)$ satisfying (\ref{parallel-transport}), the following equality holds for $\theta \in [0,s]$:
\begin{equation}\label{delta-small}
\ket{\Psi_0(\theta)} = \ket{\Phi_{\Delta}(\theta)} := U_{\Delta}(\theta) \ket{\Psi_0(0)}.
\end{equation}
\end{lemma}
\begin{proof}
The above equality follows from considering the norm of the vector
$\ket{\delta(\theta)}=\ket{\Phi_{\Delta}(\theta)}-\ket{\Psi_0(\theta)}$,
which satisfies $\ket{\delta(0)}=0$ and
$
\partial_{\theta} \ket{\delta(\theta)} = i \, \Sa(\theta) \ket{\delta(\theta)} + i \left(\Sa(\theta) - \mathcal{T}(\theta)\right)\,\ket{\Psi_0(\theta)},
$
where:
\be
\Sa(\theta)=\Sa(H(\theta),\partial_{\theta} H(\theta)), \quad
i\, \mathcal{T}(\theta) = - \frac{\one - P_0(\theta)}{H(\theta)-E_0(\theta)}\, \partial_{\theta} H(\theta),
\ee
and the definition of $i\, \mathcal{T}(\theta)$ follows from (\ref{gs:evol}).
We note that for $\Psi_n(\theta)$ an eigenstate of $H(\theta)$ with eigenvalue $E_n(\theta)$, $n \ge 0$, we have:
\begin{eqnarray}
\braket{\Psi_n(\theta)}{i\Sa(\theta) \Psi_0(\theta)} &=& \left(i\,\int_{-\infty}^{\infty} W_{\Delta}(t)\, e^{it(E_n(\theta)-E_0(\theta))} dt\right) \braket{\Psi_n(\theta)}{\partial_{\theta} H(\theta) \Psi_0(\theta)} \nonumber\\
&=& i\,\hat{W}_{\Delta}\left(E_n(\theta)-E_0(\theta)\right) \braket{\Psi_n(\theta)}{\partial_{\theta} H(\theta) \Psi_0(\theta)}\\
&=& -\frac{\braket{\Psi_n(\theta)}{\partial_{\theta} H(\theta) \Psi_0(\theta)}}{E_n(\theta)-E_0(\theta)}\quad n\ge 1,\\
\braket{\Psi_0(\theta)}{\Sa(\theta) \Psi_0(\theta)} &=& \left(\int_{-\infty}^{\infty} W_{\Delta}(t) \,dt\right) \braket{\Psi_0(\theta)}{\partial_{\theta} H(\theta) \Psi_0(\theta)} = 0,
\end{eqnarray}
where we used $\hat{W}_{\Delta}(E_n(\theta)-E_0(\theta))=i/(E_n(\theta)-E_0(\theta))$ (since $E_n(\theta)-E_0(\theta) \ge \Delta$, for $n\ge 1$, by assumption) in the third equality and $W_{\Delta}(t)$ being an odd function, for the last equality. Now, setting 
$
\ket{\Psi'(\theta)} = i(\Sa(\theta)-\mathcal{T}(\theta))\ket{\Psi_0(\theta)},
$
we have that:
\begin{eqnarray}
\braket{\Psi_n(\theta)}{\Psi'(\theta)} &=& -\frac{\braket{\Psi_n(\theta)}{\partial_{\theta} H(\theta) \Psi_0(\theta)}}{E_n(\theta)-E_0(\theta)}+\frac{\braket{\Psi_n(\theta)}{\partial_{\theta} H(\theta) \Psi_0(\theta)}}{E_n(\theta)-E_0(\theta)} = 0, \quad n \ge 1,\\
\braket{\Psi_0(\theta)}{\Psi'(\theta)} &=& 0.
\end{eqnarray}
In other words, we have shown that $\partial_{\theta} \ket{\delta(\theta)} = i \, \Sa(\theta) \ket{\delta(\theta)}$, which implies that $\ket{\delta(\theta)} = U_{\Delta}(\theta) \ket{\delta(0)} = 0$, for the unitary $U_{\Delta}(\theta)$ defined in~(\ref{eq:dynamics}). This completes the proof.
\end{proof}

\section{Quasi-adiabatic unitaries and loop operators}\label{sec:loop_unitaries}
Since we will be looking at quasi-adiabatic evolutions corresponding to Hamiltonians $H(\theta_x,\theta_y)$ with domain embedded in $\R^2$ (specifically, $(\theta_x,\theta_y) \in 2\pi \times 2\pi$), we introduce here the unitaries describing the quasi-adiabatic evolution of $\ket{\Psi_0}=\ket{\Psi_0(0,0)}$ around a closed loop in flux-space.
We begin by defining below the generators of the quasi-adiabatic dynamics: 
\begin{eqnarray}\label{def:approx_x}
\DX(\theta_x,\theta_y) &=& \Sa(H(\theta_x,\theta_y),\; \partial_{\theta_x} H(\theta_x,\theta_y))\\
\DY(\theta_x,\theta_y) &=& \Sa(H(\theta_x,\theta_y),\; \partial_{\theta_y} H(\theta_x,\theta_y)).\label{def:approx_y}
\end{eqnarray}
We continue with the unitaries corresponding to evolution through changes in the $\theta_x$ and $\theta_y$ fluxes, respectively:
\begin{eqnarray}\label{def:unitary_X}
\partial_{r} U_{X}(\theta_x,\theta_y,r) &=& i \, \DX(\theta_x+r,\theta_y)\, U_{X}(\theta_x,\theta_y,r),\quad U_{X}(\theta_x,\theta_y,0) = \one\\
\partial_{r} U_{Y}(\theta_x,\theta_y,r) &=& i \, \DY(\theta_x,\theta_y+r)\, U_{Y}(\theta_x,\theta_y,r),\quad U_{Y}(\theta_x,\theta_y,0) = \one.\label{def:unitary_Y}
\end{eqnarray}
It will be useful to note here the following composition rule for $U_X(\theta_x,\theta_y,r)$ and $U_Y(\theta_x,\theta_y,r)$:
\begin{eqnarray}\label{composition}
U_X(\theta_x+r,\theta_y,s) \, U_X(\theta_x,\theta_y,r) &=& U_X(\theta_x,\theta_y,r+s),\\
U_Y(\theta_x,\theta_y+r,s) \, U_Y(\theta_x,\theta_y,r) &=& U_Y(\theta_x,\theta_y,r+s),
\end{eqnarray}
which is easily verified upon differentiating both sides with respect to $s$ and is equivalent to evolving for ``time" $r+s$ in the $X$ and $Y$ directions, respectively, by evolving first for ``time'' $r$ and then for ``time'' $s$.

Using the above unitaries, we construct the following useful evolution operators:
\begin{eqnarray}
V[(\theta_x,\theta_y)\rightarrow (\phi_x,\phi_y)] &=& U_X(\theta_x,\phi_y,\phi_x-\theta_x)\, U_Y(\theta_x,\theta_y,\phi_y-\theta_y)\label{def:evol_ur}\\
W[(\theta_x,\theta_y)\rightarrow (\phi_x,\phi_y)] &=& U_Y(\phi_x,\theta_y,\phi_y-\theta_y)\, U_X(\theta_x,\theta_y,\phi_x-\theta_x)\label{def:evol_ru}\\
V_{\circlearrowleft}(\theta_x, \theta_y ,r) &=& V^{\dagger}[(\theta_x,\theta_y)\rightarrow(\theta_x+r, \theta_y+r)]\,W[(\theta_x,\theta_y)\rightarrow (\theta_x+r,\theta_y+r)] \label{def:unitary_loop_R}
\end{eqnarray}
Note that $V[(\theta_x,\theta_y)\rightarrow (\phi_x,\phi_y)]$ can be thought of as evolving a state along the path $\Gamma_V:(\theta_x,\theta_y) \rightarrow (\theta_x,\phi_y) \rightarrow (\phi_x,\phi_y)$ in parameter space, while $W[(\theta_x,\theta_y)\rightarrow (\phi_x,\phi_y)]$ would evolve a state along the path $\Gamma_W:(\theta_x,\theta_y) \rightarrow (\phi_x,\theta_y) \rightarrow (\phi_x,\phi_y)$. Finally, $V_{\circlearrowleft}(\theta_x, \theta_y ,r)$ is equivalent to evolving counter-clockwise around a square of side $r$ and origin $(\theta_x,\theta_y)$.

We introduce now the following important family of states:
\begin{eqnarray}\label{def:loop-states}
\ket{\Psi_{\circlearrowleft}(r)} = V_{\circlearrowleft}(0, 0 , r) \ket{\Psi_0(0,0)} = U^{\dagger}_{Y}(0, 0 , r)\,U^{\dagger}_{X}(0, r , r)\, U_{Y}(r , 0 , r)\, U_{X}(0, 0 , r) \ket{\Psi_0(0,0)}
\end{eqnarray}
These states describe the quasi-adiabatic evolution of the initial groundstate $\Psi_0(0,0)$ around a square of size $r$, starting at the origin in flux-space. 
The equality (\ref{delta-small}) allows us to calculate the geometric phase for an adiabatic evolution around a closed path near the origin:
\be\label{def:loop}
\Lambda(r) = (0,0)\rightarrow (r,0)\rightarrow(r,r)\rightarrow(0,r)\rightarrow(0,0),
\ee
where the arrows represent straight lines between two points in flux-space. In particular, we have the following:
\begin{lemma}\label{lemma:phase-estimate}
Assume that the two-parameter path of differentiable Hamiltonians $\{H(\theta_x,\theta_y)\}_{(\theta_x,\theta_y) \in \Lambda(r)}$ maintains a uniform lower bound $\Delta =\gamma/2$, on the spectral gap.
Then, for $\Psi_{\circlearrowleft}(r)$ defined in (\ref{def:loop-states}) and $\Psi_{0}(\theta_x,\theta_y)$ the groundstate of $H(\theta_x,\theta_y)$, the following phase estimate holds:
\be\label{phase-estimate}
\braket{\Psi_{0}(0,0)}{\Psi_{\circlearrowleft}(r)} = e^{i\phi(r)},
\ee
where
\begin{equation}
\phi(r) = 2 \int_0^{r} d\theta_x \int_0^r d\theta_y \, \Im\big\lbrace \braket{\partial_{\theta_y} \Psi_0(\theta_x,\theta_y)}{\partial_{\theta_x} \Psi_0(\theta_x,\theta_y)}\big\rbrace, \label{eq:phase_1}
\end{equation}
and $\Im\{\cdot\}$ stands for the imaginary part of a complex number.
\begin{proof}
We note that:
\begin{eqnarray}
U_{X}(0, 0 , r) \ket{\Psi_0(0,0)} &=&  e^{i\phi_1(r)}\ket{\Psi_0(r,0)} ,\quad i\phi_1(r) = -\int_0^{r} \braket{\Psi_0(\theta_x,0)}{\partial_{\theta_x} \Psi_0(\theta_x,0)} d\theta_x\\
U_{Y}(r, 0 , r) \ket{\Psi_0(r,0)} &=&  e^{i\phi_2(r)}\ket{\Psi_0(r,r)} ,\quad i\phi_2(r) = -\int_0^{r} \braket{\Psi_0(r,\theta_y)}{\partial_{\theta_y} \Psi_0(r,\theta_y)} d\theta_y\\
U^{\dagger}_{X}(0, r , r) \ket{\Psi_0(r,r)} &=&  e^{i\phi_3(r)}\ket{\Psi_0(0,r)} ,\quad i\phi_3(r) = \int_0^{r} \braket{\Psi_0(\theta_x,r)}{\partial_{\theta_x} \Psi_0(\theta_x,r)} d\theta_x\\
U^{\dagger}_{Y}(0, 0 , r) \ket{\Psi_0(0,r)} &=&  e^{i\phi_4(r)}\ket{\Psi_0(0,0)} ,\quad i\phi_4(r) = \int_0^{r} \braket{\Psi_0(0,\theta_y)}{\partial_{\theta_y} \Psi_0(0,\theta_y)} d\theta_y,
\end{eqnarray}
which follows immediately from (\ref{delta-small}) and the fact that the phases $\{\phi_i(r)\}_{1\le i \le 4}$ are chosen so that the respective groundstates satisfy the parallel transport condition (\ref{parallel-transport}).
Putting everything together, we get:
\be
\braket{\Psi_0(0,0)}{\Psi_{\circlearrowleft}(r)} = e^{i\phi_1(r)+i\phi_2(r)+i\phi_3(r)+i\phi_4(r)}.
\ee
Now, from Stokes' Theorem we have:
\be
\phi(r) = \phi_1(r)+\phi_2(r)+\phi_3(r)+\phi_4(r) = 2\int_0^{r} d\theta_x \int_0^r d\theta_y \, \Im\big\lbrace \braket{\partial_{\theta_y} \Psi_0(\theta_x,\theta_y)}{\partial_{\theta_x} \Psi_0(\theta_x,\theta_y)}\big\rbrace,
\ee
which completes the proof.
\end{proof}
\end{lemma}
Note that the quantity $\Im \big \lbrace \braket{\partial_{\theta_y} \Psi_0(\theta_x,\theta_y)}{\partial_{\theta_x} \Psi_0(\theta_x,\theta_y)}\big \rbrace$ is gauge-invariant; in particular, it remains constant under phase changes $\ket{\Psi'_0(\theta_x,\theta_y)} = e^{i\,f(\theta_x,\theta_y)} \ket{\Psi_0(\theta_x,\theta_y)}$, with $f(\theta_x,\theta_y)$ any real, differentiable function of $\theta_x$ and $\theta_y$.

\subsection{Enter the Hall Conductance}
To compute the Hall Conductance, we use the Kubo formula from linear response theory as applied to the setting of a torus pierced by two solenoids carrying magnetic fluxes $\theta_x$ and $\theta_y$ in the $x$ and $y$ directions, respectively~\cite{thouless}. We denote the Hall conductance at the origin in flux-space by $\sigma_{xy}$, given by the formula:
\begin{equation}\label{def:cond_0}
\sigma_{xy} = 2 \hbar \Im \left\lbrace\braket{\partial_{\theta_y}\Psi_0(0,\theta_y)_{\theta_y=0}}{\partial_{\theta_x}\Psi_0(\theta_x,0)_{\theta_x=0}}\right\rbrace
\end{equation}
where $\Psi_0(0,\theta_y)$ and $\Psi_0(\theta_x,0)$ are the ground states of $H(0,0,\theta_y,0)$ and $H(\theta_x,0,0,0)$, respectively. We note here that the units of the flux angles $\theta_x$ and $\theta_y$ are $\hbar/e$, the unit of magnetic flux divided by $2\pi$. Since we assume that $H(\theta_x,\theta_y)$ is periodic in both fluxes with period $2\pi$, we may also write for dimensionless $\theta_x,\theta_y$:
\begin{equation}\label{def:cond}
\sigma_{xy} = 2 \Im \left\lbrace\braket{\partial \theta_y \Psi_0(0,\theta_y)_{\theta_y=0}}{\partial \theta_x\Psi_0(\theta_x,0)_{\theta_x=0}}\right\rbrace \cdot \left(2\pi \, \frac{e^2}{h}\right).
\end{equation}

We now take a closer look at the phase $\phi(r)$ accumulated during the adiabatic evolution of $\Psi_0(0,0)$ along the closed path $\Lambda(r)$ defined in (\ref{def:loop}).
\begin{lemma}\label{lemma:phi(r)}
Let $\phi(r)$ be the phase defined in (\ref{eq:phase_1}). Then, the following bound holds for $r > 0$:
\begin{eqnarray}
\label{bound:phase_partials}
\left|\frac{\phi(r)}{r^2} - \sigma_{xy}\cdot \left(2\pi \, \frac{e^2}{h}\right)^{-1}\right|
\le 
\left[\sup_{s_3\in[0,r]} \left(\frac{1}{2}|[\partial_s g(s,s)]_{s=s_3}| + \frac{r}{4!}\cdot \sup_{s_4\in[0,r]}|\partial^2_s [g(s,s_4)+g(s_4,s)]_{s=s_3}|\right)\right] \cdot r,
\end{eqnarray}
with $g(s,s') = 2\Im \left\lbrace \braket{\partial_{s'} \Psi_0(s,s')}{\partial_{s} \Psi_0(s,s')}\right\rbrace$.
\begin{proof}
The above bound follows from the Taylor expansion of $\phi(r)$ up to order $3$. We have from (\ref{eq:phase_1}) that $\phi(0) = \partial_r \phi(r)|_{r=0} = 0$ and $\partial^2_r \phi(r)|_{r=0} = 4 \Im \big\lbrace\braket{\partial_{\theta_y}\Psi_0(0,\theta_y)}{\partial_{\theta_x}\Psi_0(\theta_x,0)}_{\theta_x=\theta_y=0}\big\rbrace$. This follows from the Taylor expansion of $K(r) = \int_0^r d\theta_x \int_0^r d\theta_y \, g(\theta_x,\theta_y)$, where $g(\theta_x,\theta_y)$ is doubly differentiable for $\theta_x, \theta_y \in [0,r]$. In particular, we have 
\be
\left[\partial^3_s K(s)\right]_{s=s_3} = 3 [\partial_s g(s,s)]_{s=s_3} + \int_0^{s_3} ds_4 \, [\partial^2_s (g(s,s_4)+g(s_4,s))]_{s=s_3}
\ee
which follows from higher order partials of $\left(\partial_s K(s)\right)_{s=s_1} =  \int_0^{s_1} ds [g(s,s_1)+g(s_1,s)]$. Then, expanding around $r=0$ up to third order, we get:
\be
K(r) = g(0,0)\, r^2 + \int_0^r ds_1\, \int_0^{s_1} ds_2\, \int_0^{s_2} ds_3 \, \left[3 (\partial_s g(s,s))_{s=s_3} + \int_0^{s_3} ds_4 \, [\partial^2_s (g(s,s_4)+g(s_4,s))]_{s=s_3}\right].
\ee
The above expansion implies
\be
|K(r)/r^2 - g(0,0)| \le \left[\sup_{s_3\in[0,r]} \left(\frac{1}{2}|[\partial_s g(s,s)]_{s=s_3}| + \frac{r}{4!}\cdot \sup_{s_4\in[0,r]}|\partial^2_s [g(s,s_4)+g(s_4,s)]_{s=s_3}|\right)\right] \cdot r.
\ee
In our case, $g(s,s') = 2\Im \left\lbrace \braket{\partial_{s'} \Psi_0(s,s')}{\partial_{s} \Psi_0(s,s')}\right\rbrace$. From~(\ref{def:cond}), we see that $g(0,0) = \sigma_{xy}\cdot \left(2\pi \, \frac{e^2}{h}\right)^{-1}$, which completes the proof.
\end{proof}
\end{lemma}
\subsection{Connecting the Berry phase to the Hall conductance}
Combining the simple inequality:
$$\big|e^{i\phi_1}-e^{i\phi_2}\big| = \big|e^{i(\phi_1-\phi_2)/2}-e^{-i(\phi_1-\phi_2)/2}\big| = 2\sin|(\phi_1-\phi_2)/2| = 2\int_0^{|\phi_1-\phi_2|/2} cos \phi\, d\phi \le |\phi_1-\phi_2|,$$ with Lemmas~\ref{lemma:phase-estimate} and~\ref{lemma:phi(r)}, we get the following important bound:
\begin{prop}\label{prop:adiabatic_phase_3}
For $0 < r \le \left(16\, Q_{\max} \frac{J}{\gamma} L\right)^{-1}$, the following bound holds for a numeric constant $C > 0$:
\be\label{adiabatic_phase_3}
\left|\braket{\Psi_0}{\Psi_{\circlearrowleft}(r)}^{\left(\frac{2\pi}{r}\right)^2} -e^{2\pi \,i\,\sigma_{xy}\cdot(e^2/h)^{-1}}\right| \le C \left(Q_{\max} \frac{J}{\gamma} L\right)^3\cdot r,
\ee
\begin{proof}
We note, first, that in order to use Lemma~\ref{lemma:phase-estimate}, we must set the uniform lower bound on the spectral gap of the Hamiltonians involved equal to $\gamma/2$. Recalling that the construction of the state $\Psi_{\circlearrowleft}(r)$ involves the Hamiltonian path $\Lambda(r)$ defined in~(\ref{def:loop}), we see that Lemma~\ref{lem:gap-estimate} and (\ref{bound:rot-H_X}-\ref{bound:rot-H_Y}) imply that for $0 < r \le \left(16\, Q_{\max} \frac{J}{\gamma} L\right)^{-1}$, the spectral gap is uniformly bounded below by $\gamma/2$. The bound (\ref{adiabatic_phase_3}) then follows from Lemma~\ref{lemma:phi(r)}. In particular, in order to bound the partials $\partial_s g(s,s)$ and $\partial_s \, [g(s',s)+g(s,s')]$ in (\ref{bound:phase_partials}), we make use of the bounds (\ref{bound:partial}-\ref{bound:partial_2}) and similar bounds for the norms of partial derivatives of the form $\partial^2_s, \partial_{s'} \partial_s, \partial_{s'} \partial^2_s, \partial^3_s$ acting on $\Psi_0(s',s)$ and $\Psi_0(s,s')$. The latter bounds are effectively equivalent to bounding the norms of the above set of partial derivatives acting on $H(s',0,s,0)$ and $H(s,0,s',0)$, for which we use arguments similar to those leading to the bounds (\ref{bound:rot-H_X}-\ref{bound:rot-H_Y}).
\end{proof}
\end{prop}
\section{Quantizing the Hall Conductance} 
To show that $\sigma_{xy}$ is quantized in integer multiples of $e^2/h$, it is sufficient to prove that $e^{2\pi i\,\sigma_{xy}\cdot (e^2/h)^{-1}}$ is almost-exponentially close (in the linear size, $L$, of the lattice) to $1$, since then, $\exists \,n \in \N$ such that:
\be\label{bnd:trig}
\left|\sigma_{xy}\cdot(e^2/h)^{-1}- n\right| \le \frac{\sqrt{2}}{2\pi} \left|1- e^{2\pi \,i\,\left(\sigma_{xy}\cdot(e^2/h)^{-1}- n\right)}\right|,
\ee
which follows from assuming $\big|1-e^{i\,\theta}\big|\le 1\Leftrightarrow |1-e^{i\theta}|^2 = 2(1-\cos\theta) \le 1\Leftrightarrow |\theta| \in [0,\pi/3]$, up to integer multiples of $2\pi$. In that range, $\cos|\theta|$ is monotonically decreasing from $1$ to $1/2$. In particular, we have $$\big|1-e^{i|\theta|}\big|^2 = 2(1-\cos|\theta|) = 2\int_0^{|\theta|} d\phi \int_0^{\phi} \cos\eta \, d\eta \ge 2\int_0^{|\theta|}  {\phi} \cos\phi \,d\phi \ge |\theta|^2 \cos|\theta| \ge |\theta|^2/2.$$ Since $\big|1-e^{i|\theta|}\big|=\big|1-e^{i\theta}\big|$, we get $|\theta| \le \sqrt{2} \big|1-e^{i\theta}\big|$.

Now that we have established~(\ref{bnd:trig}), we focus on the following bound:
\be\label{main_bound}
\left|1- e^{2\pi \,i\,\sigma_{xy}\cdot(e^2/h)^{-1}}\right| \le B_1+B_2+B_3,
\ee
where:
\begin{eqnarray}
B_1 &:=& \Bigl| \braket{\Psi_0}{\Psi_{\circlearrowleft}(r)}^{\left(\frac{2\pi}{r}\right)^2} -e^{2\pi \,i\,\sigma_{xy}\cdot(e^2/h)^{-1}} \Bigr| \label{B_1},\\
B_2 &:=& \Bigl|1-\braket{\Psi_0}{\Psi_{\circlearrowleft}(2\pi)} \Bigr| \label{B_2},\\
B_3 &:=& \Bigl| \braket{\Psi_0}{\Psi_{\circlearrowleft}(2\pi)}-\braket{\Psi_0}{\Psi_{\circlearrowleft}(r)}^{\left(\frac{2\pi}{r}\right)^2} \Bigr| \label{B_3}.
\end{eqnarray}
The most straightforward bound is the one for the term $B_1$, which is already given in Proposition \ref{prop:adiabatic_phase_3}. The most technically demanding bound is the one for the term $B_3$ given in~(\ref{B_3}), which requires Lieb-Robinson bounds for the quasi-adiabatic evolutions defined in the previous section. The bound for the term $B_2$ given in~(\ref{B_2}) makes use of simpler Lieb-Robinson bounds and will be addressed below.

\section{The phase around the big loop is trivial}
The bound for the second term, $B_2$, in (\ref{main_bound}) is given as Proposition~\ref{prop:gs_evol} below. Before we derive the bound, we show certain trace inequalities and develop energy estimates that allow us to study the quasi-adiabatic evolution of $\Psi_0(0,0)$ under a closed path in parameter space, when we can no longer rely on the assumption of a uniform lower bound $\Delta >0$ for the spectral gap. Note here that we consider {\it each} of the paths 
\be\label{path_components}
\Lambda_1(2\pi): (0,0)\rightarrow (2\pi,0),\, \Lambda_2(2\pi): (2\pi,0)\rightarrow (2\pi,2\pi), \, \Lambda_3(2\pi):(2\pi,2\pi)\rightarrow (0,2\pi), \, \Lambda_4(2\pi):(0,2\pi)\rightarrow (0,0)
\ee
to be closed in parameter space, since $H_0=H(2\pi,0,0,0) = H(2\pi,0,2\pi,0)=H(0,0,2\pi,0)$, which follows directly from the Aharonov-Bohm $2\pi$-periodicity of the interactions $\Phi(Z;\theta_x,0,\theta_y,0)$.

\subsection{Partial trace approximation}
\label{pta}
We prove trace inequalities that relate the reduced density matrix of the quasi-adiabatic evolution of $\ket{\Psi_0}$, both near and far from the twists driving the evolution. The estimates after tracing out sites near the twists driving the evolution are based on locality, while the estimates after tracing out sites far from the twists driving the evolution use the virtual flux idea of \cite{hast-lsm}, by comparing the reduced density matrix of interest (with flux near the origin of $T$), to the reduced density matrix of the state whose evolution involves the extra fluxes, $\phi_x, \phi_y$, near the middle of the lattice $T$ (see Fig.~\ref{fig:interactions}).
In particular, we define the following states:
\begin{eqnarray}
\ket{\Psi_X(\theta)} &=& U_X(0,0,\theta)\,\ket{\Psi_0}, \quad \ket{\Psi_X(\theta,-\theta)} = U^{(2)}_X(0,0,\theta)\,\ket{\Psi_0}, \label{psi_X}\\ 
\ket{\Psi_Y(\theta)} &=& U_Y(0,0,\theta)\,\ket{\Psi_0}, \quad \ket{\Psi_Y(\theta,-\theta)} = U_Y^{(2)}(0,0,\theta)\,\ket{\Psi_0}\label{psi_Y},
\end{eqnarray}
where the unitaries $U_X(0,0,\theta)$, $U^{(2)}_X(0,0,\theta)$, $U_Y(0,0,\theta)$ and $U^{(2)}_Y(0,0,\theta)$ are defined in (\ref{def:unitary_X}-\ref{def:unitary_Y}), with dynamics based on the one-parameter families of Hamiltonians $\{H(\theta',0,0,0)\}_{\theta'\in [0,\theta]}$, $\{H(\theta',-\theta',0,0)\}_{\theta'\in [0,\theta]}$, $\{H(0,0,\theta',0)\}_{\theta'\in [0,\theta]}$ and $\{H(0,0,\theta',-\theta')\}_{\theta'\in [0,\theta]}$, respectively.
The density matrices corresponding to the above pure states are:
\begin{eqnarray}
\rho_X(\theta) &=& \pure{\Psi_X(\theta)},\quad \rho_X(\theta,-\theta) = \pure{\Psi_X(\theta,-\theta)}\label{rho_X},\\
\rho_Y(\theta) &=& \pure{\Psi_X(\theta)},\quad \rho_Y(\theta,-\theta) = \pure{\Psi_Y(\theta,-\theta)}\label{rho_Y}.
\end{eqnarray}
To separate the action of the two sets of fluxes ($\theta_x,\theta_y$) and ($\phi_x,\phi_y$), we define the following subsets of the lattice $T$:
\be \label{sets_omega}
{\Omega_X} = \{s \in T: |x(s)| \le L/4 \}, \quad \Omega_Y = \{s \in T: |y(s)| \le L/4 \}.
\ee
For the sets $\Omega_X$ and $\Omega_Y$ above, we introduce the following Hamiltonian restrictions: 
\begin{eqnarray}
H_{\Omega_X}(\theta) &=& \sum_{Z \subset \Omega_X} \Phi(Z;\theta,0,0,0), \quad H_{\Omega_X^c}(\theta) = \sum_{Z \not \subset \Omega_X}\Phi(Z;0,-\theta,0,0),\\ 
H_{\Omega_Y}(\theta) &=& \sum_{Z \subset \Omega_Y} \Phi(Z;0,0,\theta,0), \quad H_{\Omega_Y^c}(\theta) = \sum_{Z \not \subset \Omega_Y}\Phi(Z;0,0,0,-\theta).
\end{eqnarray}
To clarify the notation above, for any subset of interactions $Z \subset T$ we write the complement as $\overline Z$.
Then, $\Omega_X^c\, (\Omega_Y^c)$ is defined as the set of sites within distance $R$ of $\overline{\Omega_X} \, (\overline{\Omega_Y})$ so that $H_{\Omega_X^c}(\theta)$ and $H_{\Omega_Y^c}(\theta)$ are supported on $\Omega_X^c$ and $\Omega_Y^c$, respectively. Finally, we note that the full Hamiltonian is a sum of terms $\Phi(Z)$ such that each interaction is either supported on $\Omega_X\, (\Omega_Y)$, or on $\Omega_X^c\, (\Omega_Y^c)$.
Using the above restrictions and recalling the definition of $H(\theta_x,\theta_y,\phi_x,\phi_y)$ in~(\ref{def:H_S}), we have the following useful decompositions:
\begin{eqnarray}\label{omega_X_decomposition}
H(\theta,0,0,0) &=& H_{\Omega_X}(\theta) + H_{\Omega_X^c}(0), \quad H(\theta,-\theta,0,0) = H_{\Omega_X}(\theta) + H_{\Omega_X^c}(\theta),\\
H(0,0,\theta,0) &=& H_{\Omega_Y}(\theta) + H_{\Omega_Y^c}(0), \quad H(0,0,\theta,-\theta) = H_{\Omega_Y}(\theta) + H_{\Omega_Y^c}(\theta).
\label{omega_Y_decomposition}
\end{eqnarray}

\begin{lemma}[Partial trace approximations]\label{lem:partial_trace}
The following partial-trace norm inequalities hold for the states and sets defined above:
\begin{flalign*}
&\max\big\{\|\Tr_{\overline{\Omega_X^c}} (\rho_X(\theta) - \rho_X(0))\|_1,\, \|\Tr_{\overline{\Omega_Y^c}} (\rho_Y(\theta) - \rho_Y(0))\|_1\big\} \le 2\, |\theta|\, (Q_{max}J L)\, g_{\Delta}(L/4-R),\\
&\max\{\|\Tr_{\overline{\Omega_X}} \left(\rho_X(\theta) - R_X(\theta, \rho_X(0))\right)\|_1,\,
 \|\Tr_{\overline{\Omega_Y}} \left(\rho_Y(\theta) - R_Y(\theta,\rho_Y(0))\right)\|_1\} \le 6\, |\theta|\,  (Q_{max}J L)\, g_{\Delta}(L/4-R),
 \end{flalign*}
 with $g_{\Delta}(\cdot)$ the almost-exponentially decaying function defined in Lemma~\ref{lemma:local_generator}.
 \begin{proof}
We only prove the bounds for $\rho_X$, since the bounds for $\rho_Y$ follow along similar lines.
Setting $H^{(1)}(\theta)=H(\theta,0,0,0)$ and $H^{(2)}(\theta)=H(\theta,-\theta,0,0)$, we note that:
\begin{eqnarray*}
\partial_{\theta'} \rho_X(\theta') &=& i\left[\Sa\left(H^{(1)}(\theta'),\partial_{\theta'} H_{\Omega_X}(\theta')\right),\rho_X(\theta')\right], \\
\partial_{\theta'} \rho_X(\theta',-\theta') &=& i\left[\Sa\left(H^{(2)}(\theta'),\partial_{\theta'} H_{\Omega_X}(\theta')\right),\rho_X(\theta',-\theta')\right] + i\left[\Sa\left(H^{(2)}(\theta'),\partial_{\theta'} H_{\Omega_X^c}(\theta')\right),\rho_X(\theta',-\theta')\right].
\end{eqnarray*}
Furthermore, recalling the definition of $\Sanewloc{M}{\Delta}(H,A)$ given in (\ref{def:local_generator}) and using the fact that $\partial_{\theta'} H_{\Omega_X}(\theta')$ has support on a strip of width $2R$ centered on the line $x=1$, we have:
\begin{eqnarray*}
\Sanewloc{L/4-R}{\Delta}(H^{(1)}(\theta'),\partial_{\theta'} H_{\Omega_X}(\theta')) = \Sanewloc{L/4-R}{\Delta}(H^{(2)}(\theta'),\partial_{\theta'} H_{\Omega_X}(\theta'))
\end{eqnarray*}
with support on $\overline{\Omega^c_X}$.
Moreover, from~(\ref{bound:rot-H_X}) and Lemma~\ref{lemma:local_generator} with $M = L/4-R$, we get:
\begin{eqnarray*}
\big\|\Sa(H^{(1)}(\theta'),\partial_{\theta'} H_{\Omega_X}(\theta'))-\Sanewloc{L/4-R}{\Delta}(H^{(1)}(\theta'),\partial_{\theta'} H_{\Omega_X}(\theta'))\big\| &\le& (Q_{max}J L)\, g_{\Delta}(L/4-R),\\
\big\|\Sa(H^{(2)}(\theta'),\partial_{\theta'} H_{\Omega_X}(\theta'))-\Sanewloc{L/4-R}{\Delta}(H^{(1)}(\theta'),\partial_{\theta'} H_{\Omega_X}(\theta'))\big\| &\le&  (Q_{max}J L) \, g_{\Delta}(L/4-R),\\
\big\|\Sa(H^{(2)}(\theta'),\partial_{\theta'} H_{\Omega_X^c}(\theta'))-\Sanewloc{L/4-R}{\Delta}(H^{(2)}(\theta'),\partial_{\theta'} H_{\Omega_X^c}(\theta'))\big\| &\le& (Q_{max}J L)\, g_{\Delta}(L/4-R).
\end{eqnarray*}
We also note that $\Sanewloc{L/4-R}{\Delta}(H^{(2)}(\theta'),\partial_{\theta'} H_{\Omega_X^c}(\theta'))$ is supported on $\overline{\Omega_X}$, since $\partial_{\theta'} H_{\Omega_X^c}(\theta')$ has support on a strip of width $2R$ centered on the line $x=L/2+1$. Equipped with the above observations, we turn our attention to proving the upper bound on $\|\Tr_{\overline{\Omega_X^c}} (\rho_X(\theta) - \rho_X(0))\|_1$:
\begin{flalign*}
&\|\Tr_{\overline{\Omega_X^c}} (\rho_X(\theta) - \rho_X(0))\|_1 
\le \int_0^\theta \left\|\Tr_{\overline{\Omega_X^c}} [\Sa(H^{(1)}(\theta'),\partial_{\theta'} H_{\Omega_X}(\theta')),\rho_X(\theta')]\right\|_1\, d\theta' \\
&\le |\theta| \sup_{\theta'\in[0,\theta]} \big\|\Tr_{\overline{\Omega_X^c}} [\Sa(H^{(1)}(\theta'),\partial_{\theta'} H_{\Omega_X}(\theta')),\rho_X(\theta')] \big\|_1
= |\theta| \sup_{\theta'\in[0,\theta]} \sup_{\stackrel{A\in \A_{\Omega^c_X}}{\|A\|=1}} |\Tr(A\,[\Sa(H^{(1)}(\theta'),\partial_{\theta'} H_{\Omega_X}(\theta')),\rho_X(\theta')])|\\
&= |\theta| \sup_{\theta'\in[0,\theta]} \sup_{\stackrel{A\in \A_{\Omega^c_X}}{\|A\|=1}} |\Tr([A,\Sa(H^{(1)}(\theta'),\partial_{\theta'} H_{\Omega_X}(\theta'))] \,\rho_X(\theta'))|
\le  |\theta| \sup_{\theta'\in[0,\theta]} \sup_{\stackrel{A\in \A_{\Omega^c_X}}{\|A\|=1}} \|[A,\Sa(H^{(1)}(\theta'),\partial_{\theta'} H_{\Omega_X}(\theta'))]\|\\
&=  |\theta| \sup_{\theta'\in[0,\theta]} \sup_{\stackrel{A\in \A_{\Omega^c_X}}{\|A\|=1}} \|[A,\Sa(H^{(1)}(\theta'),\partial_{\theta'} H_{\Omega_X}(\theta'))-\Sanewloc{L/4-R}{\Delta}(H^{(1)}(\theta'),\partial_{\theta'} H_{\Omega_X}(\theta'))]\| \le 2\, |\theta|\, (Q_{max}J L)\, g_{\Delta}(L/4-R),
\end{flalign*}
where we used the fact that $\Sanewloc{L/4-R}{\Delta}(H^{(1)}(\theta'),\partial_{\theta'} H_{\Omega_X}(\theta'))$ is supported on the complement of $\Omega^c_X$, to get the last equality.

To prove the bound on $ \|\Tr_{\overline{\Omega_X}} \left(\rho_X(\theta) - R_X(\theta, \rho_X(0))\right)\|_1$, we define the unitary $U_{{\Omega_X}}(\theta)$ with support on the set $\overline{\Omega_X^c}$, by the differential equation:
\be
\partial_{\theta'} \, U_{{\Omega_X}}(\theta') = i\, \Sanewloc{L/4-R}{\Delta}(H^{(1)}(\theta'),\partial_{\theta'} H_{\Omega_X}(\theta'))\,U_{{\Omega_X}}(\theta'),\quad U_{{\Omega_X}}(0)= \one.
\ee
Now, we have:
\begin{flalign*}
&\|\Tr_{\overline{\Omega_X}} (\rho_X(\theta) - \rho_X(\theta,-\theta))\|_1 = \big\|\Tr_{\overline{\Omega_X}} \left[U^{\dagger}_{{\Omega_X}}(\theta) \, (\rho_X(\theta) - \rho_X(\theta,-\theta))\,U_{{\Omega_X}}(\theta)\right]\big\|_1\\
 &\le \int_0^\theta \big\| \Tr_{\overline{\Omega_X}} \left[\Sa(H^{(1)}(\theta'),\partial_{\theta'} H_{\Omega_X}(\theta'))-\Sanewloc{L/4-R}{\Delta}(H^{(1)}(\theta'),\partial_{\theta'} H_{\Omega_X}(\theta')),\rho_X(\theta')\right] \big\|_1 \, d\theta' \nonumber \\
&+ \int_0^\theta \big\| \Tr_{\overline{\Omega_X}} \left[\Sa(H^{(2)}(\theta'),\partial_{\theta'} H_{\Omega_X}(\theta'))-\Sanewloc{L/4-R}{\Delta}(H^{(1)}(\theta'),\partial_{\theta'} H_{\Omega_X}(\theta')),\rho_X(\theta',-\theta')\right] \big\|_1 d\theta' \nonumber\\
 &+ \int_0^\theta \big\| \Tr_{\overline{\Omega_X}} \left[\Sa(H^{(2)}(\theta'),\partial_{\theta'} H_{\Omega^c_X}(\theta')),\rho_X(\theta',-\theta')\right] \big\|_1\, d\theta' \\
 &\le 2|\theta| \sup_{\theta'\in [0,\theta]} \big\|\Sa(H^{(1)}(\theta'),\partial_{\theta'} H_{\Omega_X}(\theta'))-\Sanewloc{L/4-R}{\Delta}(H^{(1)}(\theta'),\partial_{\theta'} H_{\Omega_X}(\theta')) \big\|\, \nonumber \\
 &+ 2|\theta| \sup_{\theta'\in [0,\theta]} \big\|\Sa(H^{(2)}(\theta'),\partial_{\theta'} H_{\Omega_X}(\theta'))-\Sanewloc{L/4-R}{\Delta}(H^{(1)}(\theta'),\partial_{\theta'} H_{\Omega_X}(\theta')) \big\|\, \nonumber \\
 &+ 2 |\theta| \sup_{\theta'\in[0,\theta]} \big\|\Sa(H^{(2)}(\theta'),\partial_{\theta'} H_{\Omega^c_X}(\theta'))-\Sanewloc{L/4-R}{\Delta}(H^{(2)}(\theta'),\partial_{\theta'} H_{\Omega^c_X}(\theta')) \big\| \\
 &\le 6\, |\theta|\,  (Q_{max}J L)\, g_{\Delta}(L/4-R).
\end{flalign*}
Finally, noting that $\rho_X(0,0) = \rho_X(0)=P_0$ and that $\{R_X(\theta', \rho_X(0,0))\}_{\theta'\in[0,\theta]}$ is the family of groundstates corresponding to the unitarily equivalent family of Hamiltonians $\{H(\theta',-\theta',0,0)\}_{\theta'\in[0,\theta]}$ which all have the same spectral gap lower bounded by $\gamma$, we have:
\be
\|\Tr_{\overline{\Omega_X}} (\rho_X(\theta,-\theta) - R_X(\theta, \rho_X(0,0)))\|_1 \le \|\rho_X(\theta,-\theta) - R_X(\theta, \rho_X(0,0))\|_1 = 0,
\ee
by applying (\ref{delta-small}) with $\Delta = \gamma$, to the quasi-adiabatic evolution of $\rho_X(0,0)$.
Using the triangle inequality:
$$\|\Tr_{\overline{\Omega_X}} \left(\rho_X(\theta) - R_X(\theta, \rho_X(0))\right)\|_1\le \|\Tr_{\overline{\Omega_X}} (\rho_X(\theta) - \rho_X(\theta,-\theta))\|_1+\|\Tr_{\overline{\Omega_X}} (\rho_X(\theta,-\theta) - R_X(\theta, \rho_X(0,0)))\|_1$$
with the above bounds, completes the proof.
\end{proof}
\end{lemma}
\subsection{Energy estimates}
We now prove a family of energy estimates that allow us to compare the true groundstate evolution to that arising from quasi-adiabatic evolution over paths in parameter space that no longer guarantee a lower bound on the spectral gap of the Hamiltonians involved.
\begin{lemma}[Energy estimate]\label{lemma:energy}
For the states $\ket{\Psi_X(\theta)}$ and $\ket{\Psi_Y(\theta)}$ defined in (\ref{psi_X}) and (\ref{psi_Y}), respectively, the following energy estimate is true for $\Delta = \gamma/2$ and $E_0$ the groundstate energy of $H_0$:
\begin{eqnarray*}
\big|\braket{\Psi_X(\theta)}{H(\theta,0,0,0)\, \Psi_X(\theta)}-E_0\big| \le 8\, |\theta| \,(Q_{\max} J^2 L^3)\, g_{\Delta}(L/4-R),
\end{eqnarray*}
with $g_{\Delta}(\cdot)$ given in~(\ref{bnd:local_generator}). The same estimate holds for $\big|\braket{\Psi_Y(\theta)}{H(0,0,\theta,0)\, \Psi_Y(\theta)}-E_0\big|$.
\begin{proof}
We will show the bound for $\Psi_X(\theta)$, since the bound for $\Psi_Y(\theta)$ follows from a similar argument. Noting that $R_X(\theta, \rho_X(0))$ is the groundstate of $H(\theta,-\theta,0,0)$, we have:
\be
E_0(0) = \Tr (H_0 \rho_X(0)) = \Tr (H(\theta,-\theta,0,0)\, R_X(\theta, \rho_X(0))) = \Tr (H_{{\Omega_X}}(\theta) \,R_X(\theta, \rho_X(0)))+ \Tr (H_{\Omega^c_X}(0) \rho_X(0))\nonumber
\ee
where we used the unitary equivalence from (\ref{twist-anti-twist}) to get the second equality and also, 
$$\Tr (H_{\Omega^c_X}(\theta) R_X(\theta, \rho_X(0))) = \Tr (H_{\Omega^c_X}(0) \rho_X(0)).$$
Now, recalling the decompositions given in (\ref{omega_X_decomposition}) and noting that $\rho_X(0) = \pure{\Psi_0}$, we have the following estimates:
\begin{flalign*}
&\big|\braket{\Psi_X(\theta)}{H(\theta,0,0,0)\, \Psi_X(\theta)}-E_0(0)\big| = |\Tr \left(H_{\Omega_X}(\theta) (\rho_X(\theta) - R_X(\theta, \rho_X(0)))\right) + \Tr \left(H_{\Omega^c_X}(0) (\rho_X(\theta) - \rho_X(0))\right)|\\
&\le |\Tr \left(H_{\Omega_X}(\theta) (\rho_X(\theta) - R_X(\theta, \rho_X(0)))\right)| + |\Tr \left(H_{\Omega^c_X}(0) (\rho_X(\theta) - \rho_X(0))\right)| \\
&\le \|H_{\Omega_X}(\theta)\|\cdot  \left(\|\Tr_{\overline{\Omega_X}} (\rho_X(\theta) - R_X(\theta, \rho_X(0)))\|_1\right) 
+ \|H_{\Omega^c_X}(0)\|\cdot \|\Tr_{{\overline{\Omega_X^c}}} (\rho_X(\theta) - \rho_X(0))\|_1\\
&\le J\, L^2\, \left(\|\Tr_{\overline{\Omega_X}} (\rho_X(\theta) - R_X(\theta,\rho_X(0)))\|_1 + \|\Tr_{{\overline{\Omega_X^c}}} (\rho_X(\theta) - \rho_X(0))\|_1\right)
\end{flalign*}
and using Lemma \ref{lem:partial_trace}, completes the proof. \end{proof}
\end{lemma}
\subsection{The phase around the big loop.}
In order to derive the next bound, we break the evolution of $\ket{\Psi_{\circlearrowleft}(2\pi)}$ around $\Lambda(2\pi)$ into its four individual components $\{\Lambda_i(2\pi)\}_{i=1}^4$ given in (\ref{path_components}). Then, applying Lemma \ref{lemma:energy} to each one, we get:
\begin{prop}\label{prop:gs_evol}
For a numeric constant $C > 0$ and $\Delta = \gamma/2$, the following bound holds:
\be\label{gs_evol}
B_2 \equiv \big|\braket{\Psi_0}{\Psi_{\circlearrowleft}(2\pi)} - 1\big| \le (C/\gamma) \,(Q_{\max} J^2 L^3) \, g_{\Delta}(L/4-R),
\ee
with $g_{\Delta}(\cdot)$ given in~(\ref{bnd:local_generator}).
\begin{proof}
We begin by noting that the $2\pi$ periodicity of $H(\theta_x,0,\theta_y,0)$ in each angle, implies:
\begin{flalign*}
&\braket{\Psi_0}{\Psi_{\circlearrowleft}(2\pi)} = \braket{\Psi_0}{U^{\dagger}_Y(0,0,2\pi)\, U^{\dagger}_X(0,0,2\pi)\,U_Y(0,0,2\pi)\,U_X(0,0,2\pi)\Psi_0}\\
&= \braket{\Psi_Y(2\pi)}{P_0\, U^{\dagger}_X(0,0,2\pi)\,P_0 \,U_Y(0,0,2\pi)\,P_0\,\Psi_X(2\pi)}
+ \braket{\Psi_Y(2\pi)}{P_0\, U^{\dagger}_X(0,0,2\pi)\,Q_0 \,U_Y(0,0,2\pi)\,P_0\,\Psi_X(2\pi)}\nonumber\\
&+ \braket{\Psi_Y(2\pi)}{P_0\, U^{\dagger}_X(0,0,2\pi)\,U_Y(0,0,2\pi)\,\delta_X(2\pi)}
+ \braket{\delta_Y(2\pi)}{U^{\dagger}_X(0,0,2\pi)\,U_Y(0,0,2\pi)\,\Psi_X(2\pi)}\\
&= |\braket{\Psi_Y(2\pi)}{\Psi_0}|^2\, |\braket{\Psi_0}{\Psi_X(2\pi)}|^2 + \braket{\Psi_Y(2\pi)}{\Psi_0}\, \braket{\delta_X(2\pi)}{\delta_Y(2\pi)}\,\braket{\Psi_0}{\Psi_X(2\pi)} \\
&+ \braket{\Psi_Y(2\pi)}{\Psi_0}\,\braket{\Psi_0}{\Psi'_X(2\pi)}\,\braket{\delta'_Y(2\pi)}{\delta_X(2\pi)} +
\braket{\Psi_Y(2\pi)}{\Psi_0}\,\braket{\delta_X(2\pi)}{U_Y(0,0,2\pi)\,\delta_X(2\pi)} 
\nonumber\\
&+ \braket{\delta_Y(2\pi)}{\delta'_X(2\pi)}\,\braket{\Psi_0}{U_Y(0,0,2\pi)\,\Psi_X(2\pi)} + \braket{\delta_Y(2\pi)}{U^{\dagger}_X(0,0,2\pi)\,\delta_Y(2\pi)}\,\braket{\Psi_0}{\Psi_X(2\pi)} \\
&+ \braket{\delta_Y(2\pi)}{U^{\dagger}_X(0,0,2\pi)\,(1-P_0)\, U_Y(0,0,2\pi)\,\delta_X(2\pi)},
\end{flalign*}
where we inserted $\one = P_0 + Q_0$ between unitaries and set:
\begin{flalign*}
&\ket{\Psi_X(2\pi)} = U_X(0,0,2\pi)\ket{\Psi_0},\quad \ket{\Psi_Y(2\pi)}=U_Y(0,0,2\pi)\ket{\Psi_0},\\
&\ket{\Psi'_X(2\pi)} = U^{\dagger}_X(0,0,2\pi)\ket{\Psi_0},\quad \ket{\Psi'_Y(2\pi)}=U^{\dagger}_Y(0,0,2\pi)\ket{\Psi_0},\\
&\delta_X(2\pi) = Q_0 \ket{\Psi_X(2\pi)} ,\quad  \delta_Y(2\pi) = Q_0 \ket{\Psi_Y(2\pi)},\quad
\delta'_X(2\pi) = Q_0 \ket{\Psi'_X(2\pi)},\quad \delta'_Y(2\pi) = Q_0 \ket{\Psi'_Y(2\pi)}.
\end{flalign*}
The above expression for $\braket{\Psi_0}{\Psi_{\circlearrowleft}(2\pi)}$ combined with the fact that 
$$|\braket{\Psi_Y(2\pi)}{\Psi_0}|^2 = 1 - \|\delta_Y(2\pi)\|^2,\quad |\braket{\Psi_X(2\pi)}{\Psi_0}|^2 = 1 - \|\delta_X(2\pi)\|^2$$ and 
$\|\delta_Y(2\pi)\|^2=\|\delta'_Y(2\pi)\|^2$, $\|\delta_X(2\pi)\|^2=\|\delta'_X(2\pi)\|^2$, gives the bound:
\be
|\braket{\Psi_0}{\Psi_{\circlearrowleft}(2\pi)}-1| \le 2(\|\delta_Y(2\pi)\|^2+\|\delta_X(2\pi)\|^2) + 4 \|\delta_Y(2\pi)\|\,\|\delta_X(2\pi)\| + \|\delta_Y(2\pi)\|^2\,\|\delta_X(2\pi)\|^2
\ee
Finally, noting that $H(2\pi,0,0,0)=H(0,0,2\pi,0)=H_0 \ge E_0 \one+ \gamma\, Q_0$, we have
\be
\|\delta_Y(2\pi)\|^2 \le \frac{1}{\gamma}\, |\braket{\Psi_Y(2\pi)}{H_0\,\Psi_Y(2\pi)}-E_0|, \quad 
\|\delta_X(2\pi)\|^2 \le \frac{1}{\gamma}\, |\braket{\Psi_X(2\pi)}{H_0\,\Psi_X(2\pi)}-E_0|
\ee
and applying Lemma \ref{lemma:energy} with $\theta = 2\pi$ to the above inequalities completes the proof.
\end{proof}
\end{prop}

\section{The phase around the small loops is uniform}
In this section, we show that the geometric phase picked up by quasi-adiabatically evolving the groundstate around small loops in flux space is independent of the starting position, up to almost-exponentially small errors in the linear size of the system, $L$.
Before giving the detailed proof, we give a high level outline.  First, in subsection \ref{ssA} we introduce the crucial fact that quasi-adiabatic continuation obeys a Lieb-Robinson bound; roughly speaking, if we take a local operator and conjugate it by the unitary $U_X(0,\theta_y,\theta_x)$, the result is still approximately local.  This Lieb-Robinson bound holds
even though the spectral gap of the Hamiltonian may vanish along the path; the reason for this is that we have defined the quasi-adiabatic continuation using a fixed $\Delta>0$
and it is this quantity $\Delta$ that controls the locality of the quasi-adiabatic evolution.
Lemma \ref{lem:local_approximation} gives the quantitative bounds that we use later. 
With this technical result in hand, in subsection \ref{ssB} we state Proposition \ref{prop:stokes} and also lemma \ref{lem:translation}, that quasi-adiabatic evolution of the ground state
along the paths of Fig.~\ref{fig:flux_path}  gives a state almost-exponentially close to the initial groundstate, up to a phase which  is independent of $\theta_x,\theta_y$, up to almost-exponentially small errors in the system size. The proof of this lemma occupies the rest of the section.

The quasi-adiabatic evolution shown in Fig.~\ref{fig:flux_path} is given by:
$V^{\dagger}[(0,0)\rightarrow(\theta_x,\theta_y)]\,V_{\circlearrowleft}(\theta_x,\theta_y,r)\, V[(0,0)\rightarrow(\theta_x,\theta_y)].$
In subsection \ref{ssC} we give an approximation to
$V^{\dagger}[(0,0)\rightarrow(\theta_x,\theta_y)]\,A_{\Omega_0} \, V[(0,0)\rightarrow(\theta_x,\theta_y)],$ for any operator
$A_{\Omega_0}$ supported on a set $\Omega_0$ defined later (this set contains points close to the origin $x=y=0$).  We call the lemma in this section the translation lemma, as it allows us to ``translate" in flux space local operators to other values of $\theta_x,\theta_y$.
Roughly, this approximation is based on the Lieb-Robinson bounds for quasi-adiabatic continuation and on the idea of virtual fluxes in subsection \ref{pta}.
In subsection \ref{ssD} we show, using a power series expansion, that $V_{\circlearrowleft}(\theta_x,\theta_y,r)$ may be approximated by such an operator $A_{\Omega_0}$.  So, we can then combine the results in subsections \ref{ssC} and \ref{ssD} to prove lemma \ref{lem:translation}. This power series expansion is carried out to the minimum order needed to obtain a nontrivial bound; higher order expansions can be used to improve the estimates if desired.

\subsection{Localizing the evolution based on $U_X(0,\theta_y,\theta_x)$}
\label{ssA}
We begin with an important lemma, \ref{lem:local_approximation} below, relating localized versions of the unitaries $U_X(0,\theta_y,\theta_x)$ and $U_X(0,0,\theta_x)$ defined in (\ref{def:unitary_X}) with $H(\theta_x,\theta_y) = H(\theta_x,0,\theta_y,0)$.

 First, we introduce the family of unitaries $U_\Omega(\theta_x,\theta_y,\theta)$ satisfying $\forall \,\theta_x , \theta_y \in [0,2\pi]$:
\be\label{unitary_omega}
\partial_{\theta} U_\Omega(\theta_x,\theta_y,\theta) = i\, \Sanewloc{L/24}{\Delta}\left(H(\theta_x+\theta,\theta_y), \partial_{\theta} H_\Omega(\theta_x+\theta,\theta_y)\right)\,U_\Omega(\theta_x,\theta_y,\theta), \quad U_\Omega(\theta_x,\theta_y,0) =\one,
\ee
with $H_\Omega(\theta_x,\theta_y)= \sum_{Z\subset \Omega} \Phi(Z;\theta_x,0,\theta_y,0)$ and
\be\label{set_omega}
\Omega = \left\{s \in T: |y(s)| \le (5/24)\,L-R \right\}.
\ee
Note that the composition rule in (\ref{composition}) implies:
\be\label{omega_composition}
U_\Omega(\theta,\theta_y,\theta_x-\theta)\, U_\Omega(0,\theta_y,\theta) = U_\Omega(0,\theta_y,\theta_x)\implies U_\Omega(0,\theta_y,\theta)\,U^\dagger_\Omega(0,\theta_y,\theta_x) = U^\dagger_\Omega(\theta,\theta_y,\theta_x-\theta),
\ee
which combined with (\ref{unitary_omega}) gives the following generating equation:
\be\label{unitary_omega_dagger}
\partial_\theta U^\dagger_\Omega(\theta,\theta_y,\theta_x-\theta) =  i\, \Sanewloc{L/24}{\Delta}\left(H(\theta,\theta_y), \partial_{\theta} H_\Omega(\theta,\theta_y)\right)\, U^\dagger_\Omega(\theta,\theta_y,\theta_x-\theta),\quad U^\dagger_\Omega(\theta_x,\theta_y,0)=\one.
\ee
The following Hamiltonian decomposition will be useful:
\be\label{hamiltonian_decomposition}
H(\theta_x,\theta_y) = H_\Omega(\theta_x,\theta_y) + H_{\Omega^c}(\theta_x,\theta_y), \quad H_{\Omega^c}(\theta_x,\theta_y) = \sum_{Z\not \subset \Omega} \Phi(Z;\theta_x,0,\theta_y,0).
\ee
Finally, since the unitary $U_\Omega(\theta_x,\theta_y,\theta)$ acts trivially outside of the set $\Omega_Y$ defined in (\ref{sets_omega}), the following is true:
\be\label{rotation_omega}
R_Y(\theta_y, U_\Omega(\theta_x,0,\theta)) = U_\Omega(\theta_x,\theta_y,\theta).
\ee

Before giving the next lemma, we introduce some notation. Just as we have used $C$ to denote various numeric constants before, we now also use $\clr$ to denote various constants which may depend upon $J/\Delta$ and $Q_{max}$, but do not depend upon any other quantities. The subscript $qa$ indicates that these quantities $\clr$ arise from the norm of certain terms that appear in the quasi-adiabatic evolution.

\begin{lemma}[Twisting Lemma]\label{lem:local_approximation}
For $U_X(0,\theta_y,\theta_x)$, $U_X(0,0,\theta_x)$ defined in (\ref{def:unitary_X}) with $H(\theta_x,\theta_y) = H(\theta_x,0,\theta_y,0)$ and $U_\Omega(\theta_x,\theta_y,\theta)$ defined in (\ref{unitary_omega}), we have, 
 for some constant $\clr > 0$ which depends upon $J/\Delta$ and $Q_{max}$,
\begin{align}
\|U^{\dagger}_X(0,\theta_y,\theta_x) \,A_{\Omega_0}\, U_X(0,\theta_y,\theta_x) - U^\dagger_\Omega(0,\theta_y,\theta_x) \,A_{\Omega_0}\, U_\Omega(0,\theta_y,\theta_x)\| &\le&  \clr \,|\theta_x|\, \|A_{\Omega_0}\|\, (\qjl ) \, g_\Delta(L/24),
\end{align}
for all $\theta_x,\theta_y \in [0,2\pi]$ and $A_{\Omega_0} \in \A_{\Omega_0}$ with 
\be\label{omega_0}
\Omega_0 = \big\{s \in T: |x(s)| \le L/8-R \mbox{ and } |y(s)| \le L/8-R\big\}.
\ee
Moreover, for all $\theta_x,\theta_y \in [0,2\pi]$ the evolved operator $U^\dagger_\Omega(0,\theta_y,\theta_x) \,A_{\Omega_0}\, U_\Omega(0,\theta_y,\theta_x)$ has support strictly within the set $\Omega_X \cap \Omega_Y$, where $\Omega_X$ and $\Omega_Y$ are defined in~(\ref{sets_omega}).
\begin{proof}
Set
$\Delta_U(\theta_x,\theta_y) \equiv U^{\dagger}_X(0,\theta_y,\theta_x) \,A_{\Omega_0}\, U_X(0,\theta_y,\theta_x)-U^\dagger_\Omega(0,\theta_y,\theta_x) \,A_{\Omega_0}\, U_\Omega(0,\theta_y,\theta_x).
$
Then, (\ref{unitary_omega_dagger}) implies:
\bea
\Delta_U(\theta_x,\theta_y) &=& \int_0^{\theta_x} \partial_{\theta} \left(U^{\dagger}_X(0,\theta_y,\theta)\,U^\dagger_\Omega(\theta,\theta_y,\theta_x-\theta) \,A_{\Omega_0}\,U_\Omega(\theta,\theta_y,\theta_x-\theta) \,U_X(0,\theta_y,\theta)\right)\, d\theta \\
 &=& i\, \int_0^{\theta_x} U^{\dagger}_X(0,\theta_y,\theta)\left[U^\dagger_\Omega(\theta,\theta_y,\theta_x-\theta) \,A_{\Omega_0}\,U_\Omega(\theta,\theta_y,\theta_x-\theta), \Delta_S(\theta,\theta_y)\right] \,U_X(0,\theta_y,\theta)\, d\theta, \label{delta_u}
\eea
where $\Delta_S(\theta,\theta_y) \equiv \Sa(H(\theta,\theta_y), \partial_{\theta} H(\theta,\theta_y))-\Sanewloc{L/24}{\Delta}(H(\theta,\theta_y), \partial_{\theta} H_\Omega(\theta,\theta_y))$.
Setting
\begin{eqnarray*}
\Delta^{(L/24)}_\Omega(\theta,\theta_y) &\equiv& \Sa(H(\theta,\theta_y), \partial_{\theta} H_{\Omega}(\theta,\theta_y)) -\Sanewloc{L/24}{\Delta}(H(\theta,\theta_y), \partial_{\theta} H_\Omega(\theta,\theta_y)) \\
\Delta^{(L/24)}_{\Omega^c}(\theta,\theta_y) &\equiv& \Sa(H(\theta,\theta_y), \partial_{\theta} H_{\Omega^c}(\theta,\theta_y))-\Sanewloc{L/24}{\Delta}(H(\theta,\theta_y), \partial_{\theta} H_{\Omega^c}(\theta,\theta_y)),
\end{eqnarray*}
we can also write:
\be
\Delta_S(\theta,\theta_y) = \Delta^{(L/24)}_\Omega(\theta,\theta_y) +\Delta^{(L/24)}_{\Omega^c}(\theta,\theta_y) + \Sanewloc{L/24}{\Delta}(H(\theta,\theta_y), \partial_{\theta} H_{\Omega^c}(\theta,\theta_y)).
\ee

Now, set 
\be\label{alpha_omega}
\alpha_\theta^\Omega(A_{\Omega_0}) \equiv U^\dagger_\Omega(\theta,\theta_y,\theta_x-\theta) \,A_{\Omega_0}\, U_\Omega(\theta,\theta_y,\theta_x-\theta)
\ee
so that $\alpha_\theta^\Omega(\cdot)$ is a quasi-adiabatic evolution based on restrictions of the differentiable family of local Hamiltonians $\{H(\theta_x,\theta_y)\}$, as can be seen from (\ref{unitary_omega_dagger}).

We turn to bounding the norm of $\Delta_U(\theta_x,\theta_y)$ starting with a triangle inequality on (\ref{delta_u}):
\begin{eqnarray}
\|\Delta_U(\theta_x,\theta_y)\| &\le&  \int_0^{\theta_x} \|[\alpha_\theta^\Omega(A_{\Omega_0}), \Delta_S(\theta,\theta_y)]\|\, d\theta \le |\theta_x| \sup_{\theta \in [0,\theta_x]} \|[\alpha_\theta^\Omega(A_{\Omega_0}), \Delta_S(\theta,\theta_y)]\|\nonumber\\
&\le& 2 |\theta_x| \|A_{\Omega_0}\| \sup_{\theta \in [0,\theta_x]} \left\{\|\Delta^{(L/24)}_\Omega(\theta,\theta_y)\|+\|\Delta^{(L/24)}_{\Omega^c}(\theta,\theta_y)\|\right\} \nonumber\\
&+& |\theta_x| \sup_{\theta \in [0,\theta_x]} \|[\alpha_\theta^\Omega(A_{\Omega_0}), \Sanewloc{L/24}{\Delta}(H(\theta,\theta_y), \partial_{\theta} H_{\Omega^c}(\theta,\theta_y))]\|\label{bound:twist}
\end{eqnarray}
From (\ref{def:local_generator}), (\ref{set_omega}), (\ref{hamiltonian_decomposition}) and (\ref{omega_0}) it follows that the generator $\Sanewloc{L/24}{\Delta}(H(\theta,\theta_y), \partial_{\theta} H_{\Omega^c}(\theta,\theta_y))$ is supported on two squares of size $L/24$ centered at $x=0,y=\pm (5/24)\,L$, separating its support from the square $\Omega_0$ by a distance $L/24$. For convenience, we will denote the support of $\Sanewloc{L/24}{\Delta}(H(\theta,\theta_y), \partial_{\theta} H_{\Omega^c}(\theta,\theta_y))$ by $\Omega_S$.

Applying the Lieb-Robinson bound derived for the quasi-adiabatic evolution $\alpha_\theta^\Omega(\cdot)$ in Thm. 4.5 in Ref.~\onlinecite{BMNS:2011} and Lem. 18 in Ref.~\onlinecite{hast-quasi}, we claim that for $\theta \in [0,2\pi]$ and a constant $\clr > 0$:
\be\label{bound:lr_twist}
\|[\alpha_\theta^\Omega(A_{\Omega_0}), \Sanewloc{L/24}{\Delta}(H(\theta,\theta_y), \partial_{\theta} H_{\Omega^c}(\theta,\theta_y))]\| \le (\clr/\Delta)\, \|A_{\Omega_0}\|\, (Q_{\max}\, J \,L\ln^2 L) \, g_\Delta(L/24),
\ee
where we applied (\ref{bound:rot-H_X}) and (\ref{def:local_generator}), to bound
$\|\Sanewloc{L/24}{\Delta}(H(\theta,\theta_y), \partial_{\theta} H_{\Omega^c}(\theta,\theta_y))\| \le (K/\Delta) \,Q_{\max}\, J \,L$.
The constant $\clr$ arises from the exponential prefactor in the Lieb-Robinson bound and may be exponentially large in $J/\Delta,Q_{max}$.
To show this result, using the notation of Section $4$ in Ref.~\onlinecite{BMNS:2011} and recalling the definitions of $f_\Delta(\cdot)$ and $g_\Delta(\cdot)$ in Lemma \ref{lem:local_decomposition} and (\ref{def:local_generator}) respectively, we have for $F_{\Psi}(d(x,y)) = (1+d(x,y))^{-3}\, f_{\Delta}(d(x,y)-R)$ and $s_0 \in \Omega_{\partial S} = \{s\in \Omega_S: d(s,\Omega_0)=L/24\}$: 
\bea
\sum_{x\in \Omega_0}\sum_{y\in \Omega_S} F_{\Psi}(d(x,y)) &\le&  |\Omega_{\partial S}|\, \sum_{d \ge 0} \sum_{r \ge L/24+d} \left|[b_{s_0}(r)\setminus b_{s_0}(r-1)]\cap \Omega_0\right| \, F_{\Psi}(r)\\
&\le& (L/12) \sum_{d\ge 0} \sum_{r \ge L/24+d} (2r) (1+r)^{-3} f_\Delta(r-R)\\
&\le& (L/6) \sum_{d\ge 0} (L/24+d)^{-1} g_\Delta(L/24+d) \sim \ln^2 L\cdot g_\Delta(L/24).
\eea
Finally, using Lemma~\ref{lemma:local_generator}, we have the bound:
\be\label{bound:local_twist}
\sup_{\theta \in [0,\theta_x]} \left\{\|\Delta^{(L/24)}_\Omega(\theta,\theta_y)\|+\|\Delta^{(L/24)}_{\Omega^c}(\theta,\theta_y)\|\right\} \le (Q_{\max}\, J \,L) \, g_\Delta(L/24).
\ee
Putting everything together in (\ref{bound:twist}), we get the desired bound:
\be
\|\Delta_U(\theta_x,\theta_y)\| \le\clr \,|\theta_x|\, \|A_{\Omega_0}\|\, (Q_{\max}\, J \,L\ln^2 L) \, g_\Delta(L/24).
\ee
\end{proof}
\end{lemma}
\subsection{Decomposing flux-space.}
\label{ssB}
Now that we have the estimates from Proposition~\ref{prop:adiabatic_phase_3} and Proposition~\ref{prop:gs_evol}, to prove the main theorem it remains to show that the following bound holds, which is proven by combining Lemma~\ref{lem:translation} below with \eqref{bound:p_1} in Appendix~\ref{append:small_loops}:
\begin{prop}\label{prop:stokes} For some $\clr > 0$, the following bound holds:
\be\label{eq:stokes}
B_3 \equiv
\big|\braket{\Psi_0}{\Psi_{\circlearrowleft}(2\pi)}-\braket{\Psi_0}{\Psi_{\circlearrowleft}(r)}^{\left(\frac{2\pi}{r}\right)^2}\big| \le \clr \left(\left(Q_{\max}\, (J/\Delta)\, L\right)^{5/2} r^{1/2} + \left(\qjl \right)^{1/2} \sqrt{g_\Delta(L/48)}\cdot r^{-1/2}\right),
\ee
Choosing
\be
\label{rchoice}
r=2\pi \left(\left\lfloor \frac{Q_{\max}\, (J/\Delta)\, L }{(\qjl)^{1/2} \sqrt{g_\Delta(L/48)}} \right\rfloor\right)^{-1}
\ee
to minimize this gives:
\be
\label{B3bound}
B_3 \leq \clr \sqrt{\ln(L)}  \left(Q_{max}(J/\Delta) \, L \right)^{3/2} \left(\Delta \cdot g_{\Delta}(L/48)\right)^{1/4},
\ee
with $g_{\Delta}(\cdot)$ the almost-exponentially decaying function defined in Lemma~\ref{lemma:local_generator}.
\end{prop}
We proceed with the proof of this bound by turning our focus on a decomposition process that breaks the large evolution around flux-space into evolutions around small loops on the $(2\pi) \times (2\pi)$ lattice. 
\begin{figure}
\centering
\includegraphics[width=280px]{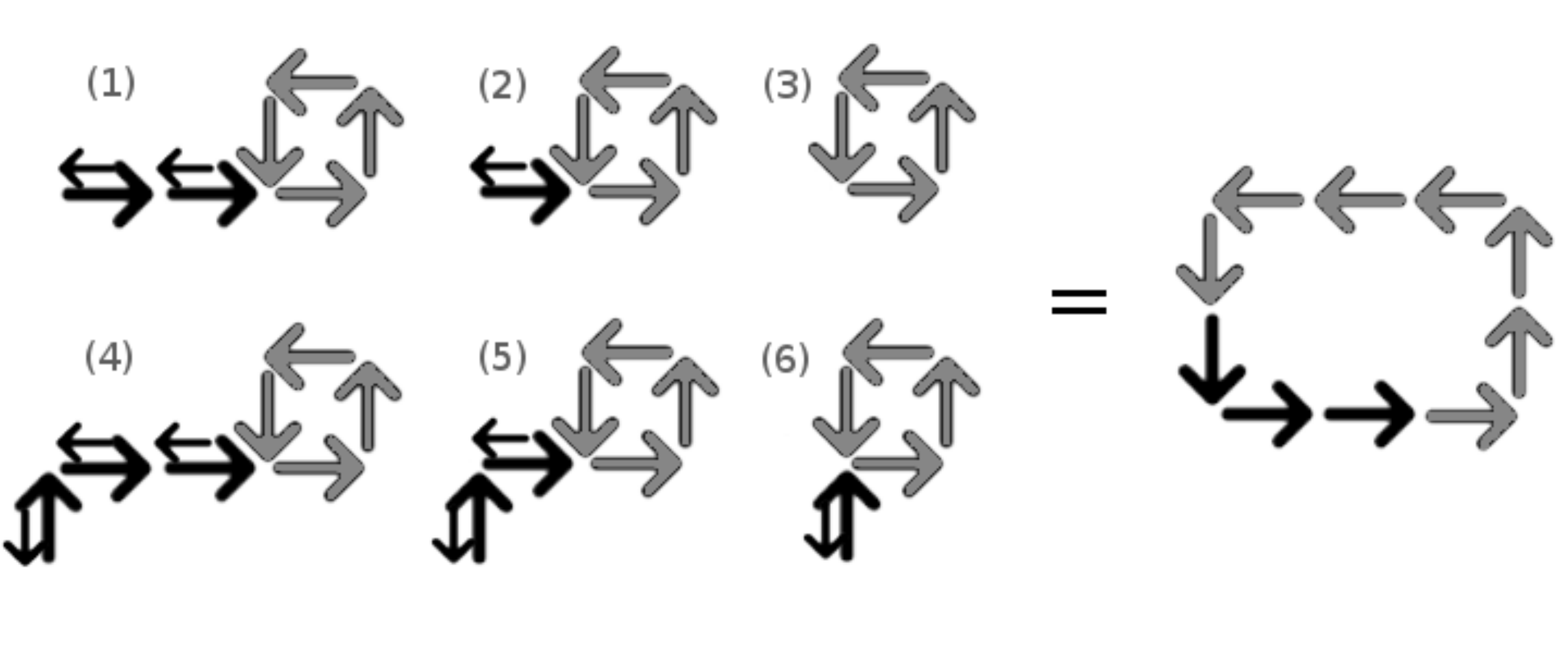}
\caption{{\small{The total evolution around a rectangle of dimension $2\times 3$ is decomposed into $6$ evolutions that all share the common features of evolving from the origin, first with $\theta_y$ and then with $\theta_x$, to reach $(\theta_x,\theta_y)$ in flux-space, then following a small counter-clockwise loop and finally, reversing the path to go from $(\theta_x,\theta_y)$ to the origin. Any $m \times n$ evolution can be decomposed in this manner, by completing first the bottom row, as in steps $(1)-(3)$, and then stacking the remaining rows on top, as in steps $(4)-(6)$.  That is, the unitary corresponding to quasi-adiabatic evolution around the larger loops is exactly equal to the
product of unitaries corresponding to evolution around the smaller loops.
}}}
\label{fig:decomposition}
\end{figure}
Fig.~\ref{fig:decomposition} describes the process used to decompose the evolution around any $m \times n$ rectangle into $m\cdot n$ evolutions involving the $r\times r$ squares which form the lattice in flux-space. This process is effectively a discrete version of Stokes' Theorem.

We define the family of states whose evolution follows the paths that appear in the decomposition process given by Fig.~\ref{fig:decomposition}:
\begin{equation}\label{def:states}
\ket{\Psi_{\circlearrowleft}(\theta_x,\theta_y,r)} = V^{\dagger}[(0,0)\rightarrow(\theta_x,\theta_y)]\,V_{\circlearrowleft}(\theta_x,\theta_y,r)\, V[(0,0)\rightarrow(\theta_x,\theta_y)]  \ket{\Psi_0}.
\end{equation}

To see why these states are important, note that projecting onto the ground-state after every individual cyclic evolution in the decomposition of $\braket{\Psi_0}{\Psi_{\circlearrowleft}(2\pi)}$ given in Fig.~\ref{fig:decomposition}, corresponds to the product of the following $(2\pi/r)^2$ terms:
\begin{equation}\label{eq:decomposition}
\braket{\Psi_0}{\Psi_{\circlearrowleft}(0,2\pi-r,r)}\braket{\Psi_0}{\Psi_{\circlearrowleft}(r,2\pi-r,r)}\cdots\braket{\Psi_0}{\Psi_{\circlearrowleft}(2\pi-r,0,r)}.
\end{equation}
The contribution from terms projecting off the ground-state is a bit more involved. It is shown in Appendix~\ref{append:small_loops} that in order to prove the bound (\ref{eq:stokes}) in Proposition~\ref{prop:stokes}, it suffices to combine Lemma \ref{lemma:phase-estimate} with the following bound proven below:
\begin{lemma}\label{lem:translation}
For sufficiently large $L$, we have $\forall \, \theta_x,\theta_y \in [0,2\pi]$:
\begin{equation}\label{eq:translation}
|\braket{\Psi_0}{\Psi_{\circlearrowleft}(\theta_x,\theta_y,r)} -\braket{\Psi_0}{\Psi_{\circlearrowleft}(r)}| \le
\clr \left(\left(Q_{\max}\, (J/\Delta)\, L\right)^5 r^5 + \left(\qjl \right) r\cdot g_\Delta(L/48)\right).
\end{equation}
\end{lemma}
Note that the above bound allows us to translate the coordinates $(\theta_x,\theta_y)$ to the origin $(0,0)$ in flux-space, where we can make use of the gap $\gamma >0$, to get good bounds on the approximation of the evolution of the groundstate $\ket{\Psi_0}$.

\subsection{The Translation Lemma}
\label{ssC}
We now focus our attention on proving (\ref{eq:translation}), but before we start we prove the following important Lemma.
\begin{lemma}\label{lem:translation_unitary}
Let $A_{\Omega_0} \in \A_{\Omega_0}$, with $\Omega_0$ defined in $(\ref{omega_0})$. Then, the following bound holds for some $\clr>0$ and $L \ge 5R$:
\begin{eqnarray*}
\big|\braket{\Psi_0}{A_{\Omega_0} \Psi_0}- \braket{\Psi_0}{V^{\dagger}[(0,0)\rightarrow(\theta_x,\theta_y)]\,R_Y(\theta_y,R_X(\theta_x,A_{\Omega_0}))\, V[(0,0)\rightarrow(\theta_x,\theta_y)]  \Psi_0}\big|\\
\le \clr (|\theta_x|+|\theta_y|) \,\|A_{\Omega_0}\| \left(\qjl \right) g_\Delta(L/24).
\end{eqnarray*}
\begin{proof}
We recall the density matrices $\rho_X(\theta_x)$ and $\rho_Y(\theta_y)$, defined in $(\ref{rho_X}-\ref{rho_Y})$. Using the rotation identity (\ref{rotation_omega}) for the auxiliary unitary $U_{\Omega}(\theta_x,\theta_y,\theta_x)$ defined in $(\ref{unitary_omega})$ and recalling the definition of $V[(0,0)\rightarrow(\theta_x,\theta_y)]$ in (\ref{def:evol_ur}), we get the following equalities:
\begin{flalign}
&\braket{\Psi_0}{V^{\dagger}[(0,0)\rightarrow(\theta_x,\theta_y)]\,R_Y(\theta_y,R_X(\theta_x,A_{\Omega_0}))\, V[(0,0)\rightarrow(\theta_x,\theta_y)]  \Psi_0} - \braket{\Psi_0}{A_{\Omega_0} \Psi_0}\nonumber\\ 
&= \Tr\left(\rho_Y(\theta_y) \, U^{\dagger}_X(0,\theta_y,\theta_x) \,R_Y(\theta_y,R_X(\theta_x, A_{\Omega_0})) \,U_X(0,\theta_y,\theta_x)\right) - \Tr(P_0 A_{\Omega_0}) \nonumber\\
&= \Tr\left(\rho_Y(\theta_y) \, \left[U^{\dagger}_X(0,\theta_y,\theta_x) \,R_Y(\theta_y,R_X(\theta_x, A_{\Omega_0})) \,U_X(0,\theta_y,\theta_x) - U^\dagger_\Omega(0,\theta_y,\theta_x) \,R_Y(\theta_y,R_X(\theta_x, A_{\Omega_0})) \,U_\Omega(0,\theta_y,\theta_x)\right]\right) \nonumber\\
&+ \Tr\left(\left[\rho_Y(\theta_y) - R_Y(\theta_y,P_0)\right] \,U^\dagger_\Omega(0,\theta_y,\theta_x) \,R_Y(\theta_y,R_X(\theta_x, A_{\Omega_0})) \,U_\Omega(0,\theta_y,\theta_x)\right) \nonumber\\
&+ \Tr\left(R_Y(\theta_y,P_0) \, R_Y\left(\theta_y, \left[U^\dagger_\Omega(0,0,\theta_x) \,R_X(\theta_x, A_{\Omega_0}) \,U_\Omega(0,0,\theta_x)-U^{\dagger}_X(0,0,\theta_x) \,R_X(\theta_x, A_{\Omega_0}) \,U_X(0,0,\theta_x)\right]\right)\right) \nonumber\\
&+\Tr\left([\rho_X(\theta_x)- R_X(\theta_x,P_0)] R_X(\theta_x, A_{\Omega_0})\right)
\end{flalign}
Finally, recalling the sets $\Omega_X$ and $\Omega_Y$ defined in (\ref{sets_omega}) and the comment preceding their definition, we have:
\begin{flalign}
&\big|\braket{\Psi_0}{V^{\dagger}[(0,0)\rightarrow(\theta_x,\theta_y)]\,R_Y(\theta_y,R_X(\theta_x,A_{\Omega_0}))\, V[(0,0)\rightarrow(\theta_x,\theta_y)]  \Psi_0} - \braket{\Psi_0}{A_{\Omega_0} \Psi_0}\big|\nonumber\\ 
&\le \big\|U^{\dagger}_X(0,\theta_y,\theta_x) \,R_Y(\theta_y,R_X(\theta_x, A_{\Omega_0})) \,U_X(0,\theta_y,\theta_x) - U^\dagger_\Omega(0,\theta_y,\theta_x) \,R_Y(\theta_y,R_X(\theta_x, A_{\Omega_0})) \,U_\Omega(0,\theta_y,\theta_x)\big\| \nonumber\\
&+ \big\|U^\dagger_\Omega(0,0,\theta_x) \,R_X(\theta_x, A_{\Omega_0}) \,U_\Omega(0,0,\theta_x)-U^{\dagger}_X(0,0,\theta_x) \,R_X(\theta_x, A_{\Omega_0}) \,U_X(0,0,\theta_x)\big\| \nonumber\\
&+ \|A_{\Omega_0}\| \left(\|\Tr_{\overline{\Omega_Y}}(\rho_Y(\theta_y) - R_Y(\theta_y,P_0))\|_1+\|\Tr_{\overline{\Omega_X}}(\rho_X(\theta_x) - R_X(\theta_x,P_0))\|_1 \right) \nonumber\\
&\le \clr \,|\theta_x| \|A_{\Omega_0}\| \left(Q_{\max}\, J \, L\ln^2 L\right) g_\Delta(L/24) + C\,(|\theta_x|+|\theta_y|) \|A_{\Omega_0}\| \left(Q_{\max}\, J \, L\right) g_\Delta(L/4-R)\label{bound:long_path}
\end{flalign}
where we used Lemmas \ref{lem:local_approximation} and \ref{lem:partial_trace}  for the two terms above. Noting that $g_\Delta(L/24) \ge g_\Delta(L/4-R)$ for sufficiently large $L$, we get the desired bound.
\end{proof}
\end{lemma}

\subsection{Localizing the loop unitary $V_{\circlearrowleft}(\theta_x,\theta_y,r)$}\label{subsec:loop_unitary}
\label{ssD}
In this section, we show that the operator $V_{\circlearrowleft}(0,0,r)$ defined in (\ref{def:unitary_loop_R}) can be approximated by the sum of operators with support on a cross of radius $L/8$ centered at the origin $x=y=0$. Moreover, we show that we may approximate $V_{\circlearrowleft}(\theta_x,\theta_y,r)$ by $R_Y(\theta_y, R_X(\theta_x, V_{\circlearrowleft}(0,0,r)))$, up to a rapidly-decaying error in $L$ and $r$.
\begin{lemma}\label{lem:local_loop_unitary}
For sufficiently large $L$, there exists a constant $C > 0$, such that for all $\theta_x,\theta_y \in [0,2\pi]$ we have the bound:
\be
\|V_{\circlearrowleft}(\theta_x,\theta_y,r)- R_Y(\theta_y, R_X(\theta_x, V_{\circlearrowleft}(0,0,r)))\| \le C\,\left(Q_{\max} (J/\Delta)\, L\right) \cdot r\cdot \left(\Delta \cdot g_\Delta(L/48) + \left(Q_{\max} (J/\Delta)\, L\right)^4 r^4\right).
\ee
Moreover, there exists an operator $W(r) \in \A_{\Omega_0}$, such that $\|W(r)-\one\| \le C \left(Q_{\max} (J/\Delta) L\right)^2 r^2$ and:
\be
\|V_{\circlearrowleft}(0,0,r) - W(r)\| \le C\,\left(Q_{\max} (J/\Delta) \, L\right) \cdot r\cdot \left(\Delta \cdot g_\Delta(L/48) + \left(Q_{\max} (J/\Delta)\, L\right)^4 r^4\right).
\ee
\end{lemma}
Before proving the above Lemma, we introduce the unitaries $U_{X(M)}(\theta_x,\theta_y,r)$ and $U_{Y(M)}(\theta_x,\theta_y,r)$ defined by the following differential equations:
\begin{eqnarray*}
\partial_r U_{X(M)}(\theta_x,\theta_y,r) &=& i\, \Sanewloc{M}{\Delta}\left(H(\theta_x+r,0,\theta_y,0), \partial_r H(\theta_x+r,0,\theta_y,0)\right) \, U_{X(M)}(\theta_x,\theta_y,r), \quad U_{X(M)}(\theta_x,\theta_y,0) = \one, \\
\partial_r U_{Y(M)}(\theta_x,\theta_y,r) &=& i\, \Sanewloc{M}{\Delta}\left(H(\theta_x,0,\theta_y+r,0), \partial_r H(\theta_x,0,\theta_y+r,0)\right) \, U_{Y(M)}(\theta_x,\theta_y,r), \quad U_{Y(M)}(\theta_x,\theta_y,0) = \one,
\end{eqnarray*}
with $\Sanewloc{M}{\Delta}(H,A)$ defined in (\ref{def:local_generator}). Later we will pick $M=L/48$ but for now $M$ is left arbitrary.  The following Lemma gives us a bound on the error of approximating the unitaries $U_X(\theta_x,\theta_y,r)$ and $U_Y(\theta_x,\theta_y,r)$ with $U_{X(M)}(\theta_x,\theta_y,r)$ and $U_{Y(M)}(\theta_x,\theta_y,r)$, respectively.
\begin{lemma}\label{lem:local_unitaries}
The following bounds hold for all $\theta_x,\theta_y \in [0,2\pi]$ and $r \ge 0$, with $M \ge 8\, R$:
\begin{eqnarray}
\|U_X(\theta_x,\theta_y,r) - U_{X(M)}(\theta_x,\theta_y,r)\| &\le& r\cdot (Q_{\max}\,J\,L)\, g_\Delta(M), \\ 
\|U_Y(\theta_x,\theta_y,r) - U_{Y(M)}(\theta_x,\theta_y,r)\| &\le& r\cdot (Q_{\max}\,J\,L)\, g_\Delta(M).
\end{eqnarray}
\begin{proof}
We only prove the bound for $U_X(\theta_x,\theta_y,r)$, since the bound for $U_Y(\theta_x,\theta_y,r)$ follows from a similar argument. First, note that $\|U_X(\theta_x,\theta_y,r) - U_{X(M)}(\theta_x,\theta_y,r)\| = \|U^{\dagger}_X(\theta_x,\theta_y,r) U_{X(M)}(\theta_x,\theta_y,r) - \one\|$.
Moreover, setting $$\Delta_M(\theta_x,\theta_y, r) = \Sanewloc{M}{\Delta}(H(\theta_x+r,0,\theta_y,0), \partial_r H(\theta_x+r,0,\theta_y,0))-\Sa(H(\theta_x+r,0,\theta_y,0), \partial_r H(\theta_x+r,0,\theta_y,0)),$$
differentiating with respect to $r$ gives:
\be
\partial_r \left(U^{\dagger}_X(\theta_x,\theta_y,r) U_{X(M)}(\theta_x,\theta_y,r)\right)_{r=s} = i\, \left(U^{\dagger}_X(\theta_x,\theta_y,s) \, \Delta_M(\theta_x,\theta_y, s)\, U_{X(M)}(\theta_x,\theta_y,s)\right)
\ee
and hence using a triangle inequality and the unitary invariance of the norm $\|\cdot\|$, we have:
\be
\|U^{\dagger}_X(\theta_x,\theta_y,r) U_{X(M)}(\theta_x,\theta_y,r) - \one\| \le \int_0^r \|\Delta_M(\theta_x,\theta_y,s)\| \, ds \le r\cdot (Q_{\max}\,J\,L)\, g_\Delta(M),
\ee
where we used \eqref{bnd:local_generator} to get the final inequality, which completes the proof.
\end{proof}
\end{lemma}
We now return to the proof of Lemma \ref{lem:local_loop_unitary}:
\begin{proof}[Proof of Lemma \ref{lem:local_loop_unitary}]
Using Lemma \ref{lem:local_unitaries} and four triangle inequalities, we may approximate the unitary defined in (\ref{def:unitary_loop_R}) as
$V_{\circlearrowleft}(\theta_x,\theta_y,r)= U_Y^{\dagger}(\theta_x,\theta_y,r)\, U_X^{\dagger}(\theta_x,\theta_y+r,r) \, U_Y(\theta_x+r,\theta_y,r) \,U_X(\theta_x,\theta_y,r)$,
with the following version:
$$V^{(M)}_{\circlearrowleft}(\theta_x,\theta_y,r)= U_{Y(M)}^{\dagger}(\theta_x,\theta_y,r)\, U_{X(M)}^{\dagger}(\theta_x,\theta_y+r,r) \, U_{Y(M)}(\theta_x+r,\theta_y,r) \,U_{X(M)}(\theta_x,\theta_y,r),$$
up to an almost-exponentially small error.
More formally, we have for all $\theta_x,\theta_y \in [0,2\pi]$:
\be\label{loop_approximation}
\big\|V_{\circlearrowleft}(\theta_x,\theta_y,r) - V^{(M)}_{\circlearrowleft}(\theta_x,\theta_y,r)\big\| \le 4\,r \cdot (Q_{\max}\,J\,L)\, g_\Delta(M).
\ee
We introduce now the following two unitaries that will aid us in the perturbative expansion of $V^{(M)}_{\circlearrowleft}(\theta_x,\theta_y,r)$:
\begin{eqnarray*}
F_r(\theta_x,\theta_y,s_1,s_2) &=& U_{Y(M)}^{\dagger}(\theta_x,\theta_y,s_1)\, U_{X(M)}^{\dagger}(\theta_x,\theta_y+r,s_2) \, U_{Y(M)}(\theta_x+r,\theta_y,s_1) \,U_{X(M)}(\theta_x,\theta_y,s_2),\\
G_r(\theta_x,\theta_y,s_1,s_2) &=& U_{Y(M)}^{\dagger}(\theta_x+r,\theta_y,s_1)\, U_{X(M)}^{\dagger}(\theta_x,\theta_y+r,s_2) \, U_{Y(M)}(\theta_x+r,\theta_y,s_1) \,U_{X(M)}(\theta_x,\theta_y+r,s_2),
\end{eqnarray*}
where $0\le s_1, s_2 \le r$.
Note that $V^{(M)}_{\circlearrowleft}(\theta_x,\theta_y,r) = F_r(\theta_x,\theta_y,r,r)$ and that the following identities hold:
\begin{eqnarray}
F_r(\theta_x,\theta_y,s_1,s_2) &=& F_r(\theta_x,\theta_y,s_1,0) \, G_r(\theta_x,\theta_y,s_1,s_2) \, F_r(\theta_x,\theta_y,0,s_2),\label{eq:F_r}\\ 
F_r(\theta_x,\theta_y,0,0) &=& G_r(\theta_x,\theta_y,0,s) = G_r(\theta_x,\theta_y,s,0) = \one, \nonumber
\end{eqnarray}
Furthermore, we define the following localized versions of the generators defined in \eqref{def:approx_x} and \eqref{def:approx_y} with $H(\theta_x,\theta_y) = H(\theta_x,0,\theta_y,0)$, recalling the decomposition given in \eqref{def:local_generator_H}: 
\be
\D^{(M)}_X(\theta_x,\theta_y) \equiv \Sanewloc{M}{\Delta}(H(\theta_x,\theta_y), \partial_{\theta_x} H(\theta_x,\theta_y)),\quad \D^{(M)}_Y(\theta_x,\theta_y) \equiv \Sanewloc{M}{\Delta}(H(\theta_x,\theta_y), \partial_{\theta_y} H(\theta_x,\theta_y)).\nonumber
\ee
In the process of showing localization for the unitary in~\eqref{eq:F_r}, the following operators will be useful:
\begin{eqnarray}
\Delta^{(M)}_X(\theta_x,\theta_y,s_2, r) &\equiv& \DX^{(M)}(\theta_x+s_2,\theta_y+r) -\DX^{(M)}(\theta_x+s_2,\theta_y)=\int_0^r \left(\partial_{s} \DX^{(M)}(\theta_x+s_2,\theta_y+s)\right) \, ds,\\
\Delta^{(M)}_Y(\theta_x,\theta_y,s_1, r) &\equiv& \DY^{(M)}(\theta_x+r,\theta_y+s_1)-\DY^{(M)}(\theta_x,\theta_y+s_1) =\int_0^r \left(\partial_{s} \DY^{(M)}(\theta_x+s,\theta_y+s_1)\right) \, ds.\label{defn:Delta_Y}
\end{eqnarray}
Note that the support of $\Delta^{(M)}_X(\theta_x,\theta_y,s_2, r)$ lies strictly within a $2M \times 4M$ box, centered at the origin in the $x-y$ orientation (i.e. $[-M,M]\times[-2M,2M]$), whereas  the support of $\Delta^{(M)}_Y(\theta_x,\theta_y,s_2, r)$ lies strictly within a $4M \times 2M$ box, centered at the origin. This follows from noting that each partial derivative eliminates terms that do not depend on the variable of differentiation, keeping in mind that interaction terms composing the above operators have been trimmed to have a radius of support $M$, thus becoming independent of $\theta_x$ and $\theta_y$ outside the box $[-M,M]\times[-M,M]$. 

In Appendix~\ref{append:expansion} we show that up to order $r^4$, for $M=L/48$ the unitaries $F_r(\theta_x,\theta_y,s_1,0)$, $G_r(\theta_x,\theta_y,s_1,s_2)$ and $F_r(\theta_x,\theta_y,0,s_2)$ are localized within the set $\Omega_0$ defined in (\ref{omega_0}). 
This result is simply the result of a Taylor expansion up to the given order; at higher order in $r$ the result would still be localized but it would be localized within some larger set.
Moreover, we show that the localized versions of the above unitaries are simple rotations of the same operators evaluated at $\theta_x=\theta_y=0$.
In particular, we have for $0\le s_1, s_2 \le r$ and a constant $C >0$, the following bounds:
\begin{eqnarray}\label{bound:F_r_1}
\|F_r(\theta_x,\theta_y,s_1,0) - R_Y(\theta_y,R_X(\theta_x, F_r(0,0,s_1,0)))\| &\le& C \left(Q_{\max} \,(J/\Delta)\, L\right)^5 r^5,\\
\|F_r(\theta_x,\theta_y,0,s_2) - R_Y(\theta_y,R_X(\theta_x, F_r(0,0,0,s_2)))\| &\le& C \left(Q_{\max} \,(J/\Delta)\, L\right)^5 r^5,\label{bound:F_r_2}\\
\|G_r(\theta_x,\theta_y,s_1,s_2) - R_Y(\theta_y,R_X(\theta_x, G_r(0,0,s_1,s_2)))\| &\le& C \left(Q_{\max} \, (J/\Delta)\, L\right)^5 r^5.\label{bound:G_r}
\end{eqnarray}

Using (\ref{loop_approximation}) with $V^{(M)}_{\circlearrowleft}(\theta_x,\theta_y,r) = F_r(\theta_x,\theta_y,r,r)$, we get from (\ref{eq:F_r}-\ref{bound:G_r}) and several triangle inequalities:
\be
\|V_{\circlearrowleft}(\theta_x,\theta_y,r)- R_Y(\theta_y, R_X(\theta_x, V_{\circlearrowleft}(0,0,r)))\| \le C\,\left(Q_{\max} (J/\Delta)\, L\right) \cdot r\cdot \left(\Delta \cdot g_\Delta(L/48) + \left(Q_{\max} (J/\Delta)\, L\right)^4 r^4\right).
\nonumber
\ee
Moreover, setting $W(r)$ to be the sum of terms up to order $4$ in $r$ in the Taylor expansion of the operator $V^{L/48}_{\circlearrowleft}(0,0,r)$ and noting that $W(0)= \one$ and $\partial_r W(r)_{r=0} = 0$, we have $\|W(r)-\one\| \le C \left(Q_{\max} (J/\Delta) L\right)^2 r^2$, for some constant $C >0$, which completes the proof of this Lemma.
\end{proof}

Now, we come back to the proof of Lemma \ref{lem:translation}, which implies Proposition~\ref{prop:stokes}. 
\begin{proof}[Proof of Lemma \ref{lem:translation}]
Using Lemmas \ref{lem:translation_unitary} and \ref{lem:local_loop_unitary} with $A_{\Omega_0} = W(r)-\one$, we get the bound (\ref{eq:stokes}) from the following estimate:
\begin{flalign*}
&|\braket{\Psi_0}{\Psi_{\circlearrowleft}(\theta_x,\theta_y,r)} -\braket{\Psi_0}{\Psi_{\circlearrowleft}(r)}| \le
|\braket{\Psi_0}{W(r) \Psi_0} - \braket{\Psi_0}{\Psi_{\circlearrowleft}(r)}|\\
&+\big|\braket{\Psi_0}{W(r) \Psi_0}- \braket{\Psi_0}{V^{\dagger}[(0,0)\rightarrow(\theta_x,\theta_y)]\,R_Y(\theta_y,R_X(\theta_x,W(r)))\, V[(0,0)\rightarrow(\theta_x,\theta_y)]  \Psi_0}\big|\\
&+\big|\braket{\Psi_0}{V^{\dagger}[(0,0)\rightarrow(\theta_x,\theta_y)]\,R_Y(\theta_y,R_X(\theta_x,[W(r)-V_{\circlearrowleft}(0,0,r)]))\, V[(0,0)\rightarrow(\theta_x,\theta_y)]  \Psi_0}\big|\\
&+\big|\braket{\Psi_0}{V^{\dagger}[(0,0)\rightarrow(\theta_x,\theta_y)]\,\left[V_{\circlearrowleft}(\theta_x,\theta_y,r)-R_Y(\theta_y,R_X(\theta_x,V_{\circlearrowleft}(0,0,r)))\right]\, V[(0,0)\rightarrow(\theta_x,\theta_y)]  \Psi_0}\big| \\
&\le 2\|W(r) -V_{\circlearrowleft}(0,0,r)\| + \|V_{\circlearrowleft}(\theta_x,\theta_y,r)-R_Y(\theta_y,R_X(\theta_x,V_{\circlearrowleft}(0,0,r)))\|\\
&+\big|\braket{\Psi_0}{(W(r)-\one) \Psi_0}- \braket{\Psi_0}{V^{\dagger}[(0,0)\rightarrow(\theta_x,\theta_y)]\,R_Y(\theta_y,R_X(\theta_x,(W(r)-\one)))\, V[(0,0)\rightarrow(\theta_x,\theta_y)]  \Psi_0}\big|,
\end{flalign*}
recalling the bound $\|W(r)-\one\| \le C \left(Q_{\max} \,(J/\Delta) L\right)^2 r^2$, for some constant $C >0$.
\end{proof}

\section{Proof of main theorem}
At this point, we can put everything together to prove the main result, the Quantization of the Hall Conductance:
\begin{proof}[Proof of the Main Theorem]
Going back to (\ref{main_bound}), and choosing $r$ to be almost-exponentially small by ~\eqref{rchoice}, we use the bounds derived in Propositions~\ref{prop:adiabatic_phase_3},~\ref{prop:gs_evol} and~\ref{prop:stokes}, to complete the proof, noting that all terms $B_1,B_2,B_3$ are almost-exponentially decaying in $L$ and the term $B_3$ dominates the terms $B_1,B_2$ for large $L$.
\end{proof}

\section{Discussion and extensions}
We have presented a proof of the quantization of Hall conductance for a system with a spectral gap on the torus.  Various extensions can be considered.  Since our goal in this paper was to present a detailed proof of the quantization of the integer Hall conductance, we only mention the extensions very briefly. 

One extension is to the fractional Hall effect; in this case, we would consider a system with a degenerate or almost degenerate groundstate subspace and a gap to the rest of the spectrum. Suppose there are $q$ groundstates for some integer $q$. To prove quantization of Hall conductance, one also needs an assumption of topological order, namely that the different groundstates are (up to small error) locally indistinguishable on sets of diameter sufficiently small compared to the linear size $L$. Under this assumption, one uses a similar proof to the integer case to show that quasi-adiabatic evolution around a small loop leaves the groundstate subspace approximately invariant, and then one uses the assumption of topological order to show that in fact the result is close to a scalar in the groundstate subspace. The new ingredient is that quasi-adiabatic evolution from $(\theta_x,\theta_y)=(0,0)$ to $(2\pi,0)$ need not be close to the identity in the groundstate subspace; rather, it is close to some general $q\times q$ unitary $u_x$. Similarly, evolution from $(0,0)$ to $(0,2\pi)$ is close to some $q\times q$ unitary $u_y$.  Thus the combined action on the ground state subspace for evolution around the large loop is close to $u_y^\dagger u_x^\dagger u_y u_x$. Using the fact that this evolution can be built up from the evolution around small loops, we can establish that $u_y^\dagger u_x^\dagger u_y u_x$ is close to a scalar, and using the fact that the determinant of this $q\times q$ unitary is one, this scalar is a $q$-th root of unity.  So, we find then that the Hall conductance is approximately quantized to an integer multiple of $(1/q)\, e^2/h$.  The calculation of the error becomes more elaborate in the fractional case, as one must also account for the small error appearing in the topological order assumption, but largely the proof is the same.

Another extension is to systems with a mobility gap but no spectral gap, or to systems on an annulus with gapless states near the edge. In the first case, one needs to give a definition of many-body localization to define the notion of a mobility gap~\cite{hast-quasi}, while in the second case one must define what it means to have a gapless edge but a gap in the bulk. We do not discuss these cases further.

{\bf Acknowledgments:} MBH thanks M. Freedman, C. Nayak, and T. Osborne for useful discussions.
SM thanks B. Nachtergaele for useful discussions on Lieb-Robinson bounds. SM acknowledges funding provided by the Institute for Quantum Information and Matter, an NSF Physics Frontiers Center with support of the Gordon and Betty Moore Foundation through Grant \#GBMF1250 and by the AFOSR Grant \#FA8750-12-2-0308.

\appendix
\begin{center}
    {\bf APPENDIX}
  \end{center}
\section{Definition of Hilbert Space of System and of Fermionic Parity}
\label{Hilbert}
We give here the definitions of the Hilbert space of the whole system and of the algebra of observables and of the fermionic parity.  These definitions are standard and are only included for completeness.
For each site $s$, we define a finite dimensional Hilbert space ${\mathcal H}_s$.  This space is defined to be the tensor product of two spaces, ${\mathcal H}_s^B \otimes {\mathcal H}_s^F$, where ${\mathcal H}_s^F$ has dimension $2^{N_o(s)}$, for some integer $N_o(s)$ which physically indicates the number of orbitals on site $s$.
Let ${\mathcal H}^B=\otimes_s {\mathcal H}_s^B$ and let ${\mathcal H}^F=\otimes_s {\mathcal H}_s^F$.  The Hilbert space ${\mathcal H}$ of the system that we study is equal to
${\mathcal H}^B \otimes {\mathcal H}^F$.

We introduce creation operators $a^\dagger_{s,j}$ and corresponding annihilation operators $a_{s,j}$
acting on ${\mathcal H}^F$.  Here, $s$ ranges over all choices of sites and $1 \leq j \leq N_o(s)$.  These operators are defined to obey the canonical anticommutation relations\cite{car}, and their algebra generates the full algebra of operators on ${\mathcal H}^F$.

We say than an operator $O_B$ acting on ${\mathcal H}^B$ is supported on a set $Z$, if $O_B$ can be written as a tensor product of an arbitrary operator acting on $\otimes_{s \in Z} {\mathcal H}_s^B$ with the identity operator acting on $\otimes_{s \not \in Z} {\mathcal H}_s^B$.  We say that an operator $O_F$ acting on ${\mathcal H}^F$ is supported on a set $Z$, if it is in the algebra generated by the creation and annihilation operators
$a^\dagger_{s,j},a_{s,j}$ for $s \in Z$. Furthermore, we define such an $O_F$ to have even fermionic parity if it is a sum of products of even numbers of such creation and annihilation operators.

We say that an operator $O$ acting on ${\mathcal H}$ is supported on a set $Z$, if $O$ can be written as a sum of tensor products $O_B \otimes O_F$, with $O_F$ and $O_B$ both supported on $Z$.  We define such an operator $O$ acting on ${\mathcal H}$ to have even fermionic parity, if $O$ can be written as a sum of tensor products $O_F \otimes O_B$, with $O_F$ having even fermionic parity and $O_B$ arbitrary. Moreover, we say that the support of an operator $O$ is equal to $Z$, if $Z$ is the smallest set such that $O$ is supported on $Z$.

Note that if two operators $O,O'$ have disjoint support and both have even fermionic parity, then they commute with each other.

\section{Derivatives of generators for quasi-adiabatic evolution}\label{append:partials}
The following general formula for a differentiable family of Hamiltonians $H(\theta)$, which can be verified by differentiating both sides with respect to $t$, will be instrumental in bounding the norm of higher partials of the generators $\DX(\theta_x,\theta_y)$ and $\DY(\theta_x,\theta_y)$ defined in \eqref{def:approx_x} and \eqref{def:approx_y} respectively:
\begin{equation}\label{eq:dH}
\left(\partial_{\theta} e^{itH(\theta)}\right) e^{-itH(\theta)} =  -e^{itH(\theta)} \left(\partial_{\theta} e^{-itH(\theta)}\right) =  i \int_0^t \tau_u^{H(\theta)} \left(\partial_{\theta} H(\theta)\right) du, \quad \mbox{where} \quad \tau_u^H(A) := e^{iuH} A e^{-iuH}.
\end{equation}
Using the first equality above, it is straightforward to derive the commutator terms for the partials below:
\begin{eqnarray*}
\partial_{\theta_x} \tau_u^{H(\theta_x,\theta_y) }(\partial_{\theta_y}H(\theta_x,\theta_y)) &=&\tau_u^{H(\theta_x,\theta_y)}\left(\partial_{\theta_x} \partial_{\theta_y} \, H(\theta_x,\theta_y) \right)+
\left[\left(\partial_{\theta_x} e^{iuH(\theta_x,\theta_y) }\right) e^{-iuH(\theta_x,\theta_y)}, \tau_u^{H(\theta_x,\theta_y) }(\partial_{\theta_y}H(\theta_x,\theta_y))\right],\\
\partial_{\theta_y} \tau_u^{H(\theta_x,\theta_y) }(\partial_{\theta_y}H(\theta_x,\theta_y)) &=&\tau_u^{H(\theta_x,\theta_y)}\left(\partial^2_{\theta_y} \, H(\theta_x,\theta_y) \right)+
\left[\left(\partial_{\theta_y} e^{iuH(\theta_x,\theta_y) }\right) e^{-iuH(\theta_x,\theta_y)}, \tau_u^{H(\theta_x,\theta_y) }(\partial_{\theta_y}H(\theta_x,\theta_y))\right].
\end{eqnarray*}
From (\ref{eq:dH}) we have:
$\left(\partial_{\theta_x} e^{iuH(\theta_x,\theta_y)}\right) e^{-iuH(\theta_x,\theta_y)}
= i \int_0^u \tau_s^{H(\theta_x,\theta_y)}(\partial_{\theta_x}H(\theta_x,\theta_y)) \, ds,$
and similarly for $\partial_{\theta_y}$.
Setting $H(\theta_x,\theta_y) := H(\theta_x,0,\theta_y,0)$ and using the above formulas with the definition of $\DY(\theta_x,\theta_y)$, we get:
\begin{eqnarray*}
\big\|\partial_{\theta_x} \DY(\theta_x,\theta_y)\big\| &\le& \big\|\partial_{\theta_x} \partial_{\theta_y} \, H(\theta_x,\theta_y)\big\| \int_{-\infty}^{\infty} |t| \,W_{\Delta}(t)\, dt
+\big\|\partial_{\theta_y} \, H(\theta_x,\theta_y)\big\| \big\|\partial_{\theta_x} \, H(\theta_x,\theta_y)\big\| \int_{-\infty}^{\infty} t^2 \,W_{\Delta}(t) \, dt,
\\
\big\|\partial_{\theta_y} \DY(\theta_x,\theta_y)\big\| &\le& \big\|\partial^2_{\theta_y} \, H(\theta_x,\theta_y)\big\| \int_{-\infty}^{\infty} |t| \,W_{\Delta}(t)\, dt
+ \big\|\partial_{\theta_y} \, H(\theta_x,\theta_y)\big\|^2 \int_{-\infty}^{\infty} t^2 \,W_{\Delta}(t) \, dt,
\end{eqnarray*}
with similar bounds for partial derivatives of $\DX(\theta_x,\theta_y)$.
Now, recalling that $W_{\Delta}(t)$ is almost-exponentially decaying, we have $\int_{-\infty}^{\infty} |t| \,W_{\Delta}(t)\, dt =c_1/\Delta$ and $\int_{-\infty}^{\infty} t^2 \,W_{\Delta}(t)\, dt = c_2/\Delta^2$. Finally, using estimates similar to the ones derived in (\ref{bound:rot-H_X}-\ref{bound:rot-H_Y}), it should be clear that the above partials have norms bounded by $C\, (Q_{\max} (J/\Delta)\, L )^2$, for some $C >0$. In general, one can show that $m$-th order partials of $\DY(\theta_x,\theta_y)$ and $\DX(\theta_x,\theta_y)$ have norms bounded by $C\, (Q_{\max} (J/\Delta)\, L)^{m+1}$. The same arguments apply to partial derivatives of $\DY^{(M)}(\theta_x,\theta_y)$ and $\DX^{(M)}(\theta_x,\theta_y)$.
 
\section{Decomposing the evolution into small loops.}\label{append:small_loops}
Let $N= 2\pi/r$ and define $U_{(N-m)+nN} = V^{\dagger}(m\cdot r,n\cdot r)V_{\circlearrowleft}(m\cdot r,n\cdot r,r) V(m\cdot r,n\cdot r),$ for $m,n\in[0,N-1]$ to be the evolution operator corresponding to the cyclic path leading to and around the square with lower-left corner at $(m\cdot r,n\cdot r)$ in flux-space (see Fig.~\ref{fig:decomposition}). Then, we have the following identity:
\begin{equation}
V_{\circlearrowleft}(0,0,2\pi) = U_{N^2}\, U_{N^2-1} \cdots U_3\, U_2\, U_1
\end{equation}
as can be verified by the decomposition described in Fig.~\ref{fig:decomposition}. To help us track the contribution from each term in the decomposition as we insert $P_0=\pure{\Psi_0}$ and $Q_0=\one-P_0$ after each cyclic evolution, we introduce the scalars:
$$p_{[s,t]} = \braket{\Psi_0}{U_t\, U_{t-1}\cdots U_s \Psi_0},\qquad \mbox{and}\qquad
q_{[s,t]} = \bra{\Psi_0}\, U_t\, U_{t-1}\cdots U_{s-1} Q_0 U_s \ket{\Psi_0}.$$
Moreover, we define $p_t =p_{[t,t]}= \braket{\Psi_0}{U_t \Psi_0}$.  In particular, $p_N = \braket{\Psi_0}{\Psi_\circlearrowleft(0,0,r)} = \braket{\Psi_0}{\Psi_\circlearrowleft(r)}$.
The quantity we want to bound in (\ref{eq:stokes}) is 
$\braket{\Psi_0}{\Psi_{\circlearrowleft}(2\pi)}-\braket{\Psi_0}{\Psi_{\circlearrowleft}(r)}^{\left(\frac{2\pi}{r}\right)^2} = p_{[1,N^2]}-(p_N)^{N^2}.$
From the triangle inequality, we get:
\be
\big|p_{[1,N^2]}-(p_N)^{N^2}\big| \le \big|p_{[1,N^2]}-p_1\, p_{2}\cdots p_{N^2-1}\, p_{N^2}\big| + \big|p_1\, p_{2}\cdots p_{N^2-1}\, p_{N^2}-(p_N)^{N^2}\big|,
\ee 
so it suffices to bound each term on the right.
We start with a bound for $\big|p_{[1,N^2]}-p_1\, p_{2}\cdots p_{N^2-1}\, p_{N^2}\big| $. Using the following equalities recursively:
$
p_{[1,N^2]} = p_1 p_{[2,N^2]} + q_{[1,N^2]}, \,
p_{[2,N^2]} = p_2 p_{[3,N^2]} + q_{[2,N^2]}, \ldots,
$
we may write:
\begin{equation}\label{bound:p_0}
p_{[1,N^2]} - p_1\, p_{2}\cdots p_{N^2-1}\, p_{N^2} = \sum^{N^2-1}_{i=1} p_1\cdots p_{i-1} q_{[i,N^2]}
\end{equation}
We turn our attention to the terms $q_{[i,N^2]}$. We begin by observing that the following series of bounds hold:
\begin{eqnarray}
|q_{[i,N^2]}| &=& |\braket{\Psi_0}{U_{N^2}\cdots U_{i+1} Q_0 U_i \Psi_0}| \le \|Q_0 U_i\ket{\Psi_0}\|
= \sqrt{1 - |p_i|^2} \le \sqrt{2(|p_N| - |p_i|)} \le \sqrt{2|p_i-p_N|}.\nonumber
\end{eqnarray}
where we used $1=|p_N|$ (see Lemma \ref{lemma:phase-estimate}) and $|p_i| \le 1, \forall i \in [1,N]$.
Since, we have $N^2-1$ terms in (\ref{bound:p_0}), the triangle inequality  and the above bound imply:
\begin{equation}\label{off_gs}
\big|p_{[1,N^2]} - p_1\, p_{2}\cdots p_{N^2-1}\, p_{N^2}\big|\le (N^2-1) \sup_{i\in [1,N^2]} \sqrt{2|p_i-p_N|}.
\end{equation}
Moreover, assuming $\sup_{i\in[1,N^2]} |p_i - p_N| = \delta$ for some $\delta \in [0,2]$, we get:
\begin{equation}
\big|(p_{N})^{N^2} - p_1\, p_{2}\cdots p_{N^2-1}\, p_{N^2}\big| \le \left(1+\delta\right)^{N^2}-1 \le e^{\delta N^2}-1 = \int_0^{\delta N^2} e^y \, dy \le e^{\delta N^2} \, \delta N^2,
\end{equation}
where the first inequality follows from expanding the product on the right of the identity: 
\begin{equation*}
p_1\, p_{2}\cdots p_{N^2-1}\, p_{N^2} = (p_N + [p_1-p_N])\cdots (p_N + [p_{N^2-1}-p_N])\,(p_N + [p_{N^2}-p_N])
\end{equation*}
and then, after subtracting the term $(p_N)^{N^2}$, using the triangle inequality along with $|p_N| =1$. The second inequality follows from comparing the binomial expansion of $(1+x/m)^m$ term by term with the Taylor expansion of $e^x$ and using the simple inequality $\binom{m}{k} \le \frac{m^k}{k!}$.
Putting everything together, we get the final bound:
\be\label{bound:p_1}
\big|\braket{\Psi_0}{\Psi_{\circlearrowleft}(2\pi)}-\braket{\Psi_0}{\Psi_{\circlearrowleft}(r)}^{\left(\frac{2\pi}{r}\right)^2}\big| \le 4\pi^2 \left(\sqrt{2\, \delta\cdot r^{-4}} + e^{4\pi^2 \delta\cdot r^{-2}} \delta\cdot r^{-2} \right),
\ee
recalling that $\delta := \sup_{i\in [1,N^2]} |p_i-p_N|$ is bounded in (\ref{eq:translation}).

\section{Taylor expansions of operators $F_r(\theta_x,\theta_y,s_1,s_2)$ and $G_r(\theta_x,\theta_y,s_1,s_2)$}\label{append:expansion}
The notation used in this section is defined in Subsection~\ref{subsec:loop_unitary}.
We begin by showing localization for $F_r(\theta_x,\theta_y,s_1,0)$. A similar argument applies to $F_r(\theta_x,\theta_y,0,s_2)$. One can easily check that:
\be
\partial_s F_r(\theta_x,\theta_y,s,0)_{s=s_1} = i\, U_{Y(M)}^{\dagger}(\theta_x,\theta_y,s_1)\, \Delta^{(M)}_Y(\theta_x,\theta_y,s_1, r)\, U_{Y(M)} (\theta_x,\theta_y,s_1) \cdot F_r(\theta_x,\theta_y,s_1,0).\nonumber
\ee
Noting that we only need to consider partial derivatives up to order $3$ since $\|\Delta^{(M)}_Y(\theta_x,\theta_y,s_1, r)\| = \mathcal{O}(r)$ (see~\eqref{defn:Delta_Y} and Appendix~\ref{append:partials} for a discussion on how to bound $\|\partial_{s} \DY^{(M)}(\theta_x+s,\theta_y+s_1)\|$), we have:
\begin{eqnarray*}
\partial_s F_r(\theta_x,\theta_y,s,0)_{s=0} &=& i\,\Delta^{(M)}_Y(\theta_x,\theta_y,0, r)\\
\partial^2_s F_r(\theta_x,\theta_y,s,0)_{s=0} &=& - \left[\Delta^{(M)}_Y(\theta_x,\theta_y,0, r), D^{(M)}_Y(\theta_x,\theta_y)\right] + i\, \partial_s \Delta^{(M)}_Y(\theta_x,\theta_y,s, r)_{s=0} - \Delta^2_Y(\theta_x,\theta_y,0, r)\\
\partial^3_s F_r(\theta_x,\theta_y,s,0)_{s=0} &=& -i \left[ \left[\Delta^{(M)}_Y(\theta_x,\theta_y,0, r), D^{(M)}_Y(\theta_x,\theta_y)\right], D^{(M)}_Y(\theta_x,\theta_y)\right] \\
&-& \partial_s \left[\Delta^{(M)}_Y(\theta_x,\theta_y,s, r), D^{(M)}_Y(\theta_x,\theta_y+s)\right]_{s=0}
+ i\, \partial^2_s \Delta^{(M)}_Y(\theta_x,\theta_y,s, r)_{s=0} + \mathcal{O}(r^2).
\end{eqnarray*}
Now, the crucial point is that the operators $\Delta^{(M)}_Y(\theta_x,\theta_y,s, r)$ and $D^{(M)}_Y(\theta_x,\theta_y+s)$ satisfy for $M \le L/24$:
\begin{eqnarray*}
&&\Delta^{(M)}_Y(\theta_x,\theta_y,s, r) = R_Y(\theta_y,R_X(\theta_x,\Delta^{(M)}_Y(0,0,s, r))), \\
&&\left[\Delta^{(M)}_Y(\theta_x,\theta_y,s, r), D^{(M)}_Y(\theta_x,\theta_y+s) -R_Y\left(\theta_y,R_X\left(\theta_x,D^{(M)}_Y(0,s)\right)\right)\right] = 0,\\
&&\left[R_Y\left(\theta_y,R_X\left(\theta_x,\left[\Delta^{(M)}_Y(0,0,s, r), D^{(M)}_Y(0,s)\right)\right]\right), D^{(M)}_Y(\theta_x,\theta_y+s) -R_Y\left(\theta_y,R_X\left(\theta_x,D^{(M)}_Y(0,s)\right)\right)\right] = 0,
\end{eqnarray*}
which may be verified by considering the action of the twists $R_Y(\theta_y,\cdot)$ and $R_X(\theta_x,\cdot)$ on interaction terms with support contained in $\Omega_0 = \big\{s \in T: |x(s)| \le L/8-R \mbox{ and } |y(s)| \le L/8-R\big\}$. For example, note that for the truncated Hamiltonians defined in the statement of Lemma~\ref{lem:local_decomposition} with $M \le L/2$, we have $R_Y(\theta_y, H_{Z(M)}(0,0,s,0)) = H_{Z(M)}(0,0,\theta_y+s,0)$ for $Z$ with support near $y=1$ and $R_X(\theta_x, H_{Z(M)}(s,0,0,0)) = H_{Z(M)}(\theta_x+s,0,0,0))$ for $Z$ with support near $x=1$. This follows from the trivial action of the truncated Hamiltonians $H_{Z(M)}(0,0,s,0)$ and $H_{Z(M)}(s,0,0,0)$ near the twists at $y=L/2+1$ and $x=L/2+1$, respectively, recalling that the interaction terms  commute with charge operators whose support covers the range of the interaction (see~\eqref{defn:interaction} and~\eqref{twist-anti-twist}). In general, the previous argument can be applied to any Hamiltonian that acts trivially (i.e. has no interaction terms supported) on one of the two boundary lines of the charge operators $Q_X$ and/or $Q_Y$, effectively introducing a \emph{single} boundary twist by applying the corresponding global twist $R_X(\theta_x,\cdot)$ and/or $R_Y(\theta_y,\cdot)$ on the Hamiltonian.
\begin{figure}
\centering
\includegraphics[width=280px]{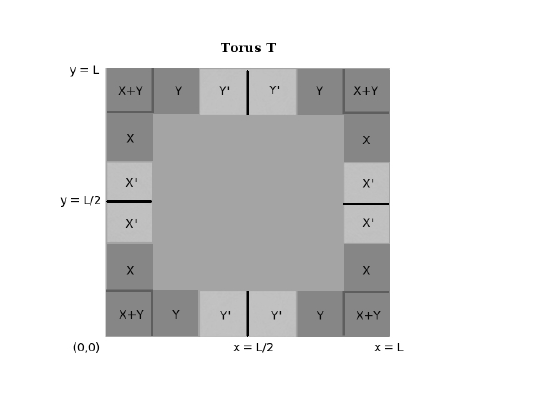}
\caption{\small{Decomposition of the torus $T$ into areas where the operators $\Sanewloc{M}{\Delta}(H(0,\theta_x,r,\theta_y),\partial_{\theta} H(\theta,0,r,\theta_y)|_{\theta=0})$ and $\Sanewloc{M}{\Delta}(H(r,\theta_x,0,\theta_y),\partial_{\theta} H(r,\theta_x,\theta,0)|_{\theta=0})$ act non-trivially. The dark shaded regions are defined as $X=\{s \in T: x(s) \in [-L/6,L/6] \, \wedge \, y(s) \in [-L/3-R, L/3+R] \}$ and $Y=\{s \in T: y(s) \in [-L/6,L/6] \, \wedge \, x(s) \in [-L/3-R, L/3+R] \}$ and correspond to interaction terms $\Sanewloc{M}{\Delta}(H(0,0,r,0),\partial_{\theta} \Phi(Z;\theta,0,r,0)|_{\theta=0})$ with $Z \subset X$ and $\Sanewloc{M}{\Delta}(H(r,0,0,0),\partial_{\theta} \Phi(Z;r,0,\theta,0)|_{\theta=0})$ with $Z \subset Y$, respectively. On the other hand, the light shaded regions $X'=\{s \in T: x(s) \in [-L/6,L/6] \, \wedge \, y(s) \in [2L/3, 4L/3] \}$ and $Y'=\{s \in T: y(s) \in [-L/6,L/6] \, \wedge \, x(s) \in [2L/3, 4L/3] \}$ correspond to terms $\Sanewloc{M}{\Delta}(H(0,0,0,\theta_y),\partial_{\theta} \Phi(Z;\theta,0,0,\theta_y)|_{\theta=0})$ with $Z \subset X'$ and $\Sanewloc{M}{\Delta}(H(0,\theta_x,0,0),\partial_{\theta} \Phi(Z;0,\theta_x,\theta,0)|_{\theta=0})$ with $Z \subset Y'$, respectively. Since each interaction has radius of support $M \le L/6$, interactions in $X'$ commute with interactions in $Y$ and $Y'$ and interactions in $Y'$ further commute with those in $X$.}}
\label{fig:space_decomp}
\end{figure}

Of course, in order for the previous argument to work, we need to show that each term we wish to rotate globally has support away from the twists at $x=L/2+1$ and $y=L/2+1$. To study the support of the terms derived in the above partials, we choose one of these terms as an illustrative example. In particular, we note that taking commutators of $\Delta^{(M)}_Y(\theta_x,\theta_y,s, r)$ with $D^{(M)}_Y(\theta_x,\theta_y+s)$, can only extend the support of $\Delta^{(M)}_Y(\theta_x,\theta_y,s, r)$ (a $4M \times 2M$ box, centered at the origin) up to $2M$ in the $y$-direction, so terms affected by the twist on $x=L/2$ and $y=L/2$ will not appear in the commutator for $M\le L/12$, since then $(2+2)M\le L/2-2M$. Finally, since the support of $D^{(M)}_Y(\theta_x,\theta_y+s) -R_Y\left(\theta_y,R_X\left(\theta_x,D^{(M)}_Y(0,s)\right)\right)$ does not intersect the support of $\Delta^{(M)}_Y(\theta_x,\theta_y,s, r)$, the commutator of the two operators is trivially $0$.

The above observations, combined with the partials we computed and the assumption $0\le s_1, s_2 \le r$, imply that for a constant $C >0$, the following bounds hold:
\begin{eqnarray}
\|F_r(\theta_x,\theta_y,s_1,0) - R_Y(\theta_y,R_X(\theta_x, F_r(0,0,s_1,0)))\| &\le& C \left(Q_{\max} \,(J/\Delta)\, L\right)^5 r^5,\\
\|F_r(\theta_x,\theta_y,0,s_2) - R_Y(\theta_y,R_X(\theta_x, F_r(0,0,0,s_2)))\| &\le& C \left(Q_{\max} \,(J/\Delta)\, L\right)^5 r^5,
\end{eqnarray}
with the constant power of $(Q_{\max} \, (J/\Delta)\, L)$ coming from bounds on the norms of fifth-order commutators containing terms like $D^{(M)}_Y(\theta_x,\theta_y+s_1)$ and partials like $\partial_{s} \DY^{(M)}(\theta_x+s,\theta_y+s_1)$, whose norms may be bounded using (\ref{naive_bnd:sa}) and arguments similar to the one found in Appendix~\ref{append:partials}, respectively.

Moreover, the supports of the partial derivatives  of $F_r(0,0,s_1,0)$ and $F_r(0,0,0,s_2)$ up to order $4$ in $r$, lie strictly within $\Omega_0$, for $M \le L/48$. By taking $M$ to be a smaller fraction of $L$, we could have continued the expansion to higher orders in $r$, while keeping the supports of $F_r(0,0,s_1,0)$ and $F_r(0,0,0,s_2)$ strictly within $\Omega_0$. For our purposes, it suffices to consider errors of order $\mathcal{O}(r^5)$.

We turn our focus, now, to showing the bound:
\begin{eqnarray}
\|G_r(\theta_x,\theta_y,s_1,s_2) - R_Y(\theta_y,R_X(\theta_x, G_r(0,0,s_1,s_2)))\| &\le& C \left(Q_{\max} \, (J/\Delta)\, L\right)^5 r^5
\end{eqnarray}
for some constant $C > 0$, while keeping the support of $G_r(0,0,s_1,s_2)$ within $\Omega_0$, up to order $4$ in $r$ (recalling that $0 \le s_1,s_2 \le r$).
We begin by setting:
\begin{eqnarray*}
\Delta^{r}_{Y}(\theta_x,\theta_y,s_1, s_2) &\equiv& i\, U_{X(M)}^{\dagger}(\theta_x,\theta_y+r,s_2)\, D^{(M)}_Y(\theta_x+r,\theta_y+s_1) \, U_{X(M)} (\theta_x,\theta_y+r,s_2) - i\, D^{(M)}_Y(\theta_x+r,\theta_y+s_1)\\  
&=& \int_0^{s_2} U_{X(M)}^{\dagger}(\theta_x,\theta_y+r,s)\left[D^{(M)}_X(\theta_x+s,\theta_y+r),D^{(M)}_Y(\theta_x+r,\theta_y+s_1)\right] U_{X(M)} (\theta_x,\theta_y+r,s) \, ds
\\
\Delta^{r}_{X}(\theta_x,\theta_y,s_1, s_2) &\equiv& i\, D^{(M)}_X(\theta_x+s_2,\theta_y+r) -i\, U_{Y(M)}^{\dagger}(\theta_x+r,\theta_y,s_1)\, D^{(M)}_X(\theta_x+s_2,\theta_y+r) \, U_{Y(M)} (\theta_x+r,\theta_y,s_1)\\ 
&=& \int_0^{s_1} U_{Y(M)}^{\dagger}(\theta_x+r,\theta_y,s)\left[D^{(M)}_X(\theta_x+s_2,\theta_y+r),D^{(M)}_Y(\theta_x+r,\theta_y+s)\right] U_{Y(M)} (\theta_x+r,\theta_y,s) \, ds
\end{eqnarray*}
Taking partials with respect to the variables $s_1$ and $s_2$, one may verify:
\begin{eqnarray*}
\partial_{s_1'} G_r(\theta_x,\theta_y,s_1',s_2)_{s_1'=s_1} &=& G_r(\theta_x,\theta_y,s_1,s_2) \cdot \left(U_{Y(M)}^{\dagger}(\theta_x+r,\theta_y,s_1)\, \Delta^{r}_{Y}(\theta_x,\theta_y,s_1, s_2)\, U_{Y(M)} (\theta_x+r,\theta_y,s_1)\right) \\
\partial_{s_2'} G_r(\theta_x,\theta_y,s_1,s_2')_{s_2'=s_2} &=& \left(U_{X(M)}^{\dagger}(\theta_x,\theta_y+r,s_2)\, \Delta^{r}_{X}(\theta_x,\theta_y,s_1, s_2)\, U_{X(M)} (\theta_x,\theta_y+r,s_2)\right) \cdot G_r(\theta_x,\theta_y,s_1,s_2)
\end{eqnarray*}
Using the above relations, it is straightforward to check:
\begin{eqnarray*}
\partial_{s_1} G_r(\theta_x,\theta_y,s_1,s_2)_{s_1=0} &=& \Delta^{r}_{Y}(\theta_x,\theta_y,0, s_2)\\
\partial_{s_2} G_r(\theta_x,\theta_y,s_1,s_2)_{s_2=0} &=& \Delta^{r}_{X}(\theta_x,\theta_y,s_1, 0)\\
\partial^2_{s_1} G_r(\theta_x,\theta_y,s_1,s_2)_{s_1=0} &=&
i\left[\Delta^{r}_{Y}(\theta_x,\theta_y,0, s_2),D^{(M)}_Y(\theta_x+r,\theta_y)\right] + \left(\partial_{s_1} \Delta^{r}_{Y}(\theta_x,\theta_y,s_1, s_2)\right)_{s_1=0} + (\Delta^{r}_{Y}(\theta_x,\theta_y,0, s_2))^2\\
\partial^2_{s_2} G_r(\theta_x,\theta_y,s_1,s_2)_{s_2=0} &=&
i\left[\Delta^{r}_{X}(\theta_x,\theta_y,s_1, 0),D^{(M)}_X(\theta_x,\theta_y+r)\right] + \left(\partial_{s_2} \Delta^{r}_{X}(\theta_x,\theta_y,s_1, s_2)\right)_{s_2=0} + (\Delta^{r}_{X}(\theta_x,\theta_y,s_1,0))^2
\end{eqnarray*}
We note here that we only need to evaluate the following partials at $s_1=s_2=0$ in order to have a complete picture of $G_r(\theta_x,\theta_y,s_1,s_2)$ up to order $4$ in $r$:
\begin{eqnarray*}
\partial_{s_2} (\partial_{s_1} G_r(\theta_x,\theta_y,s_1,s_2)_{s_1=0})_{s_2=0} &=& \partial_{s_2} \Delta^{r}_{Y}(\theta_x,\theta_y,0, s_2)_{s_2=0} = \left[D^{(M)}_X(\theta_x,\theta_y+r),D^{(M)}_Y(\theta_x+r,\theta_y)\right]\\
\partial_{s_2} \left(\partial^2_{s_1} G_r(\theta_x,\theta_y,s_1,s_2)_{s_1=0}\right)_{s_2=0} &=& i \left[\left[D^{(M)}_X(\theta_x,\theta_y+r),D^{(M)}_Y(\theta_x+r,\theta_y)\right],D^{(M)}_Y(\theta_x+r,\theta_y)\right] \\
&+& \left[D^{(M)}_X(\theta_x,\theta_y+r), \left(\partial_{s_1} D^{(M)}_Y(\theta_x+r,\theta_y+s_1)\right)_{s_1=0}\right]\\
\partial_{s_1} \left(\partial^2_{s_2} G_r(\theta_x,\theta_y,s_1,s_2)_{s_2=0}\right)_{s_1=0} &=& i \left[\left[D^{(M)}_X(\theta_x,\theta_y+r),D^{(M)}_Y(\theta_x+r,\theta_y)\right],D^{(M)}_X(\theta_x,\theta_y+r)\right] \\
&+& \left[\left(\partial_{s_2} D^{(M)}_X(\theta_x+s_2,\theta_y+r)\right)_{s_2=0}, D^{(M)}_Y(\theta_x+r,\theta_y)\right]\\
\partial^2_{s_1} \left(\partial^2_{s_2} G_r(\theta_x,\theta_y,s_1,s_2)_{s_2=0}\right)_{s_1=0}  &=& i
\left[\left(\partial^2_{s_1} \Delta^{r}_{X}(\theta_x,\theta_y,s_1, 0)\right)_{s_1=0},D^{(M)}_X(\theta_x,\theta_y+r)\right] \\
&+& \left(\partial_{s_2} \left(\partial^2_{s_1} \Delta^{r}_{X}(\theta_x,\theta_y,s_1, 0)\right)_{s_1=0}\right)_{s_2=0}
+ 2 \left[D^{(M)}_X(\theta_x,\theta_y+r),D^{(M)}_Y(\theta_x+r,\theta_y)\right]^2
\end{eqnarray*}
Now, since $\left(\partial^2_{s_1} \Delta^{r}_{X}(\theta_x,\theta_y,s_1, 0)\right)_{s_1=0}= \partial_{s_2} \left(\partial^2_{s_1} G_r(\theta_x,\theta_y,s_1,s_2)_{s_1=0}\right)_{s_2=0},$
we can see from Figure \ref{fig:space_decomp} that all of the above commutators are rotated versions of the same commutators with $\theta_x=\theta_y=0$, where we apply $R_Y(\theta_y, R_X(\theta_x, \cdot))$ to perform the rotation as in the study of $F_r(\theta_x,\theta_y,s_1,0)$. For example, we have for one of the fourth-order terms implicit in the first term of the last line of partials evaluated above:
\begin{eqnarray*}
&&\left[\left[\left[D^{(M)}_X(\theta_x,\theta_y+r),D^{(M)}_Y(\theta_x+r,\theta_y)\right],D^{(M)}_Y(\theta_x+r,\theta_y)\right], D^{(M)}_X(\theta_x,\theta_y+r)\right] = \\
&&R_Y\left(\theta_y,R_X\left(\theta_x, \left[\left[\left[D^{(M)}_X(0,r),D^{(M)}_Y(r,0)\right],D^{(M)}_Y(r,0)\right], D^{(M)}_X(0,r)\right]\right)\right)
\end{eqnarray*}
where we have assumed that $M \le L/12$ to make sure that all terms affected by the twists at $x=y=L/2$ vanish. Furthermore, choosing $M = L/48$, guarantees that up to fourth order all commutators have support within $\Omega_0$. To see this, note that each new commutator includes terms with support at most $2M$ away from the support of each term in the commutator. For example, the above commutator has, potentially, the largest support of all the terms up to fourth order. It is supported within a cross of radius $5M$, with two axis of width $2M$ each, centered at the origin.


\end{document}